\documentclass[letterpaper, 10pt, letter]{IEEEtran} 
\IEEEoverridecommandlockouts                            
\usepackage[english]{babel}
\usepackage[utf8,utf8x]{inputenc}
\usepackage{graphicx}
\usepackage[colorinlistoftodos]{todonotes}
\usepackage{hyperref}
\usepackage{listings}
\usepackage{amsmath,amscd,amssymb,amsgen,amsfonts,amsbsy}
\usepackage{mathrsfs}
\usepackage{ragged2e}
\usepackage{gensymb}
\usepackage{textcomp}
\usepackage{float}
\usepackage{latexsym}
\usepackage{mathtools}
\usepackage{breqn}
\usepackage{tikz}
\usetikzlibrary{automata,arrows,positioning,calc}
\usepackage{authblk}
\usepackage{url}
\usepackage{cite}
\usepackage{algorithmic}
\usepackage[ruled,lined]{algorithm2e}
\usepackage{breqn}
\usepackage{caption}
\usepackage{subcaption}
\usepackage{breqn}
\usepackage{tabularx}
\usepackage{amsthm}

\newtheorem{definition}{Definition}
\title{An Intent Modeling and Inference Framework for Autonomous and Remotely Piloted Aerial Systems }
\author{Kesav Kaza, Varun Mehta, Hamid Azad, Miodrag Bolic, Iraj Mantegh% 
\thanks{This work has been supported by the National Research Council Canada.}
\thanks{Kesav Kaza, Hamid Azad and Miodrag Bolic are with the School of Electrical Engineering and Computer Science, 
University of Ottawa, Canada. Varun Mehta and Iraj Mantegh are with National Research Council Canada.
}
} 

\begin{document}

\maketitle

\begin{abstract}
An intent modeling and inference framework is presented to assist the defence planning for protecting a geo-fence against unauthorized flights. First, a novel mathematical definition for the \emph{intent} of an uncrewed aircraft system (UAS) is presented. The concepts of critical waypoints and critical waypoint patterns are introduced and associated with a motion process to fully characterize an intent. This modeling framework consists of representations of UAS mission planner, used to plan the aircraft's motion sequence, as well as a defense planner, defined to protect geo-fence. It is applicable to autonomous, semi-autonomous, and piloted systems in 2D and 3D environments with obstacles. The framework is illustrated by defining a library of intents for a security application. Detection and tracking of the target are presumed for formulating the intent inference problem. Multiple formulations of the decision maker’s objective are discussed as part of a deep-learning based methodology. Further, a multi-modal dynamic model for characterizing the UAS flight is discussed. This is later utilized to extract features using the interacting multiple model (IMM) filter for training the intent classifier. Finally, as part of the simulation study, an attention-based bi-directional long short-term memory (Bi-LSTM) network for intent inference is presented. The simulation experiments illustrate various aspects of the framework including trajectory generation, radar measurement simulation, etc., in 2D and 3D environments.
\end{abstract}

%\begin{abstract}
%An intent modelling and inference framework is presented to assist the decision making process for protecting a geo-fence against unauthorized flights. First, a generic mathematical definition for \emph{intent} of an uncrewed aerial vehicle (UAV) system is presented.% along with a modeling framework to assist in the decision-making process for protecting a geo-fence against unauthorized flights.  
%The concepts of critical waypoints and critical waypoint pattern are introduced, and are associated with a motion process to fully characterize an intent. This modeling framework consists of a representation of both the UAV mission planner who plans the motion of UAV, and the defence planner who is the tasked with protecting the geo-fence. It is applicable to both autonomous, semi-autonomous and piloted systems in 2D and 3D environments with obstacles. The framework is illustrated by defining a library of intents for a security application. Detection and tracking of the target are presumed for formulating the intent inference problem. Multiple formulations of the decision maker's objective are discussed as part of a deep learning based methodology. Finally, an attention based bi-directional LSTM network for intent inference is presented as part of the simulation study.  The simulation experiments illustrate various aspects of the framework including feature extraction, radar measurement simulation, etc., in 2D and 3D environments. %and also points to important directions for future work. 
%\end{abstract}

\section{Introduction}
Uncrewed Aircraft Systems or Uncrewed Aerial Vehicles (UAVs), also commonly known as drones, have seen rapid adoption across various domains, including agriculture, environmental monitoring, disaster management, and even last-mile delivery services. Their versatility and potential for positive impact are undeniable. However, alongside their beneficial applications, UAVs also pose security challenges and potential risks.
The problem of intent recognition, prediction or inference has been widely studied from various perspectives for civilian and military aircraft \cite{Hwang2008:ProbabilisticIntentBased, Zhang:IEEEAccess2019:JointTrackingInference, balaji2013stochastic}, ground vehicles \cite{yoder2012monitoring, Wang2011:StochastiGMTI}, and human operators \cite{Moolchandani:ASME2022:IntentUnstructured}. However, this has been relatively less true for UAVs until recently\cite{Liang2021:maliciousdetection,Perrusquia2023, Perrusquia2024:PhysicsInspired, yun2023estimation,Lee2022probabilistic}. In the following, we provide a discussion of the important aspects of intent inference of aerial systems, including conventional aircraft and UAVs. 
%Table~\ref{tab:litreview} briefly summarizes the relevant literature on intent inference.

%
\begin{table*}[!htb]
\caption{Relevant Intent Inference Literature.} {\scriptsize{(Abbreviations: AoI - area of interest, RMIMM - residual mean interacting multiple model, GMTI - ground moving target indication, ADS-B - automatic dependent surveillance–broadcast, POMDP - partially observable Markov decision process, ATC - air traffic control)}} 
\label{tab:litreview}
\begin{center}
{\scriptsize{
\begin{tabularx}{\linewidth}%{|l|X X X X X X|} %{|l|>{\justify\arraybackslash}X|>{\justify\arraybackslash}X|>{\justify\arraybackslash}X|>{\justify\arraybackslash}X|>{\justify\arraybackslash}X|>{\justify\arraybackslash}X|}
%{|l|>{\raggedright\arraybackslash}X|>{\raggedright\arraybackslash}X|>{\raggedright\arraybackslash}X|>{\raggedright\arraybackslash}X|>{\raggedright\arraybackslash}X|>{\raggedright\arraybackslash}X|}
%{|l|X|X|X|X|X|X|}
{l|>{\raggedright\arraybackslash}X >{\raggedright\arraybackslash}X >{\raggedright\arraybackslash}X >{\raggedright\arraybackslash}X >{\raggedright\arraybackslash}X >{\raggedright\arraybackslash}X}
\hline
 & \textbf{Target type}  
 & \textbf{Location and sensor type}     
 & \textbf{Intent types}
 & \textbf{Measurements/Features}                     
 & \textbf{Methods used}
 & \textbf{Model/Formulation}  
 \\ \hline
 Perrusquía 2024\cite{Perrusquia2024:PhysicsInspired}
 & Drone   
 & Asset, staring radar                               
 &{Mapping, Point-to-Point, Package Delivery, Perimeter Flights}
 & Kinematic, Type of mission and expected flight path 
 & Convolutional Bi-LSTM and LSTM autoencoder
 & Trajectory prediction and classification 
 \\ \hline
 Perrusquía 2023\cite{Perrusquia2023}
 & Drone   
 & Generic sensor type with arbitrary location
 & Mission represented by objective function
 & States, control input, tracking error, obj. function values
 & Policy error inverse reinforcement learning algorithm
 & Inverse reinforcement learning of mission objective function 
 \\ \hline
 Yun 2023\cite{yun2023estimation} 
 & Drone   
 & Geo-fence, staring radar                            
 & Smuggling,Kamikaze, Surveillance
 & Trajectory(positional) frequencies 
 & Naïve Bayes classifier 
 & Classification problem
 \\ \hline
 Lee 2022\cite{Lee2022probabilistic} 
 & Commercial Aircraft 
 & Airports, lidar/camera  
 & Ground movements (towards apron, away from apron, irrelevant to apron, stop)  
 & Kinematic, beacon signals, surrounding objects, context features
 & Context-aware multi-modal sensor fusion based method
 & Probabilistic Bayesian network, Classification
 \\ \hline
 Liang 2021\cite{Liang2021:maliciousdetection}
 & Drone                                              
 & Generic sensor type                                
 & Enter, leave or remain in AoI
 & Generic target state (spatio-temporal measurements)
 & Sequential estimation of the probability of a destination in AoI
 & Latent state estimation problem
 \\ \hline
 Zhang 2019\cite{Zhang:IEEEAccess2019:JointTrackingInference}
 & Military aircraft (Cargo/fighter)
 & Generic sensor type                                 
 & Cruise or attack
 & Kinematic and target attribute senor measurements   
 & Bayesian state-intent-class joint density estimation          
 & Joint Identification and Tracking                   
 \\ \hline
 Lowe 2014\cite{Lowe:2014:Phdthesis_intentprediction}  
 & Manned Civilian aircraft   
 & ADS-B messages  
 & Behavior intents – \{comply/don’t, goal states, navigation states\}   
 & Position, velocity, acceleration, mode, trajectory change point     
 & Hybrid state estimation using RMIMM filter  
 & Hidden Markov Model, POMDP           
 \\ \hline
 Yoder 2012\cite{yoder2012monitoring} 
 & Land vehicles/human targets          
 & Imaging using small UAV  
 & Intent to enter levels of proximity  
 & Proximity, direction, visibility, offroad detection
 & Using distance measures, Threshold rule
 & Classification problem  
 \\ \hline
 Wang 2011\cite{Wang2011:StochastiGMTI}  
 & Ground targets   
 & Airborne GMTI radar    
 & Intention to maintain certain attack formations (column, wedge, pincer) 
 & GMTI measurements  
 & Spatial trajectory identification, matching geometric patterns  
 & Hidden Markov Model         
 \\ \hline
 Hwang 2008\cite{Hwang2008:ProbabilisticIntentBased} 
 & Manned Civilian Aircraft   
 & ATC radar  
 & Regulation and flight plan related intents  
 & ATC radar measurements 
 & Kalman filter based heuristic algorithm 
 & Intent inference or verification of broadcast intent 
 \\ \hline
 Yepes 2007\cite{yepes2007:trajectoryprediction} 
 & Manned Civilian Aircraft   
 & FIS broadcast, ADS-B messages 
 & Intents constrained by ATC regulations. 
 \{horizontal, vertical, speed intents\}   
 & Position, velocity, acceleration, mode, trajectory change point          
 & Hybrid state and mode estimation using RMIMM        
 & State estimation and trajectory prediction          \\ \hline
\end{tabularx}
}}
\end{center}
\end{table*}

The literature on intent inference is diverse in terms of intent representation/encoding, problem formulation and solution methodologies. 
The problem has been formulated in different ways such as classification, state estimation, trajectory prediction, identification and tracking, inverse reinforcement learning, etc. In this paper, we consider an UAV's intent from its flight trajectory and motion pattern perspective, and will closely examine a common use case of securing a geographical area against UAV intrusion.  

%Similarly, the literature can be classified based on their representation of intents. The various broad descriptions of intent include: 
Broadly, an aircraft's intent has been represented as trajectory behaviors or waypoint sequences that an aircraft is supposed to navigate. The relative relationship between an aircraft's current measured state and the intended (expected) trajectory behaviors is then used to infer the current intent (see \cite{Lowe:2014:Phdthesis_intentprediction,  Tran:2022:trajpred}). However, in the literature, there is considerable variation in both the qualitative definitions of intent and their mathematical descriptions, depending on the application context. Some of this variation among the intent representations and the underlying rationale is discussed below. In this work, we aim to present a simple and versatile mathematical model for UAV intent that captures the key aspects of existing intent models related to critical asset protection. 
%In the following, we discuss the rationale behind the various intent models in the literature.
%In this work, we aim to present a compact yet broadly applicable mathematical model for UAV intent with the ability to condense most of the model diversity in the literature on critical asset protection.

%(1) reaching a specific destination or entering an area of interest (AoI), (2) library of trajectory behaviours based on the application, (3) deviation from expected behaviour. 
Much of the variation among intent representations follows from the need to transform intent from a complex cognitive construct to a low dimensional representation as intent classes.
In simple surveillance/security applications, intent is generally inferred from whether a target UAV is aiming to enter or reach a specific destination within an area of interest \cite{Liang2021:maliciousdetection}. In this case it is a binary variable dictating whether to enter or not, and is hidden from the decision maker tasked with the security of the AoI. In more complex applications where the targets have multi-dimensional action spaces, a library of intents is defined. Some security applications require identifying multiple intents such as ``kamikaze attack'', ``smuggling'', ``image acquisiton'', etc. as in \cite{yun2023estimation} or ``mapping flights'', ``point-to-point flights'', ``package delivery'', ``perimeter flights'' as in \cite{Perrusquia2024:PhysicsInspired}. This approach is justified by the difference in the mis-classification and also interception costs associated with various intents in security applications (for example surveillance versus attack). 
Defining an intent library is also applicable to scenarios where there is structure imposed on the aircraft pilot's navigation model by air traffic regulations. In the case of civilian aircraft, the pilot's intent library can be modelled using the flight plan structure and traffic regulation knowledge. These can be further sub-categorized as intents in horizontal plane, vertical plane, speed related, etc. \cite{yepes2007:trajectoryprediction, Lowe:2014:Phdthesis_intentprediction}.
For crewed commercial civilian airlines, conflict avoidance is synonymous with conforming to air traffic control regulations. In this case, the target's conformance versus deviation from expected behaviour is a useful way of defining intent \cite{Gariel2011:trajectoryclustering}. Recently, \cite{Wang2023:swarmintention} considered expansion, free-movement and contraction as possible intents for drone swarms.
%
%
%Let us now look at the different intent models that have been used in the literature for formulating the inference problem. 
%
Whatever the model, intents are inevitably associated with flight path or motion patterns. Two important ways of describing target motion include flight-mode based models and stochastic differential equations (SDE). For both these models the motion can  finally be represented using a state space model of the dynamical system \cite{Liang2021:maliciousdetection,Perrusquia2024:PhysicsInspired}. In mode based models, the flight is assumed to be in one of a finite set of modes such as constant velocity (cruising), constant acceleration (ascent or descent), coordinated turn (maneuvering), etc. In classical aviation literature motion patterns are also characterized using mode change points and trajectory change points \cite{ Krozel2006:intentwithpath, yepes2007:trajectoryprediction, Lowe:2014:Phdthesis_intentprediction}. In case of the SDE model the motion is generated by a destination or a virtual leader as a hidden variable in a carefully designed stochastic process \cite{Liang2021:maliciousdetection, Gan2021:levyprocess}. 
Table~\ref{tab:litreview} summarizes the relevant literature on intent inference.

All the above literature use application-specific definitions of intent and work with model-specific target behaviour, with diverse models and methods. For instance, even for the same (informal) objective such as ``protection of critical assets'', several types of intent descriptors can be used (such as smuggling, mapping etc, as discussed above \cite{yun2023estimation, Perrusquia2024:PhysicsInspired}).

\begin{figure*}[!htp]
    \centering
    \includegraphics[width=1\linewidth]{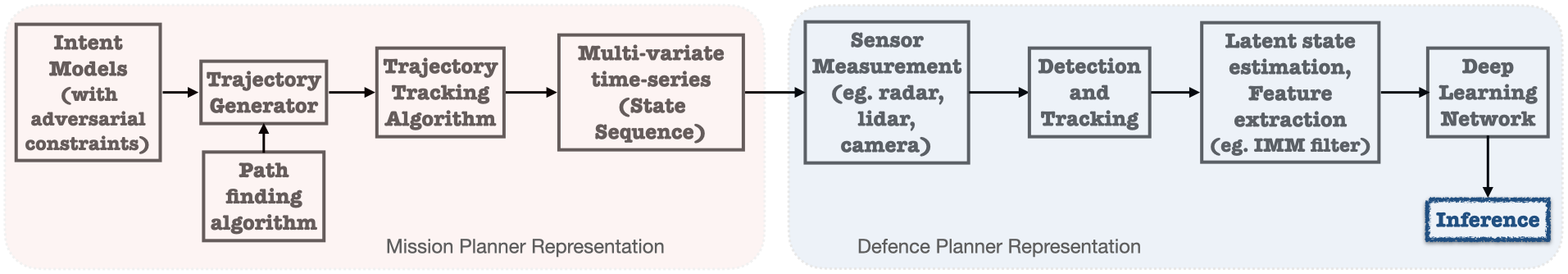} 
    \caption{The intent inference framework with a representation of the mission planner and the defence planner. The mission planner drives a UAV in the environment with a hidden intent, and the defence planner infers this intent using the specified modules. In general, the mission planner's representation is from the perspective of the defence planner.}
    \label{fig:intent-framework}
\end{figure*}

In this paper, we attempt to present an integrated mathematical framework for planning the protection of critical assets. This can be achieved through a robust quantitative characterization of intent which includes a representation of the \emph{mission planner} who plans the mission of the UAV, as well as the \emph{defence planner} who plans the protection of an asset using sensing, detection, tracking, followed by intent inference and potential interception. This representation summarized by Fig.~\ref{fig:intent-framework}.
%We propose to do this by inferring the intent of UAVs which are in proximity to a geo-fenced region, to assist the decision-maker tasked with protecting an asset, so that they might be intercepted in advance and prevented. 
%
%
The formulation of the intent inference problem presumes prior detection and tracking of the target, estimation of kinematic state information, i.e.,  position, velocity, acceleration, and turn rate. Estimation and tracking are, in general, done by algorithms that are based on variations of Kalman filter and the interacting multiple model (IMM).
Further, in most of the work mentioned above the problem formulations are mainly focused on intent inference through trajectory forecasting, destination prediction using a set of possible final destinations, and/or intent classification using trajectory datasets for training. 
For the purpose of trajectory prediction, carefully chosen motion models have been used. This does not account for the latest developments in motion planning algorithms which are commonly used for path finding and trajectory optimization of autonomous navigation systems. Our proposed framework utilizes motion/path planning algorithms (e.g.\cite{deFilippis:JIRS12:3Dpathplanning,Koenig:JAI04:LPA*,Ragi:AES13:UAV_POMDP,meuleau2010pomdp,miller2009pomdp,Tu2023:DQNpathplanning}) to generate motion patterns associated with various intents. 
%IntentInferenceFrameworkSchematic_color.png
%
%
%
%
%\subsubsection*{Brief Overview of Motion Planning algorithms} 
%Path finding and motion planning for autonomous aerial systems has been studied extensively for different missions, constraints and use cases \cite{deFilippis:JIRS12:3Dpathplanning,Koenig:JAI04:LPA*}. 
%Mission planning methodologies have been proposed in the literature based on one or more of sets of constraints. In general most methodologies have separate modules for path planning and trajectory optimization. Path planning is to find a path to the intended destination or target while trajectory optimization considers other kinematic motion constraints with additional objectives such as generating smooth trajectories with minimum energy utilization. \cite{deFilippis:JIRS12:3Dpathplanning,Koenig:JAI04:LPA*,Ragi:AES13:UAV_POMDP,meuleau2010pomdp,miller2009pomdp,Tu2023:DQNpathplanning}). 
%We utilize some path planning methodologies for the modelling and representation of intelligent UAV mission planners, and characterization of their intents. 
%
%
%As we have seen in the above discussion, intents of agents can only be meaningfully defined with reference to an environment and the UAV mission planner's objective. In this paper, we focus on the problem of approaching or breaching a geo-fenced boundary and prediction of the operator's intent in this setting. We define intent of UAVs with respect to region of interest enclosed by a geo-fence $\mathcal{G}$. Generally, the overall objective is to protect $\mathcal{G}$ from intrusion.
%
%

\subsection*{Summary of Contributions}
%\begin{itemize}
We present an integrated intent inference framework with a novel generic mathematical definition of intent. This can be used to represent a wide variety of intent types described in the literature. 

 To do this, we define the concept of critical waypoints; each intent is modeled using a critical waypoint pattern and an associated motion process. The motion process can be described using a motion planning algorithm. 

The major advantage of this framework is that critical waypoint regions can be specified by an expert pilot or by mission planning algorithms or be derived from flight experiment. Hence, the framework is applicable for intent inference in both manually planned and autonomous missions. This also allows for the possibility of using hybrid datasets with real and synthetic data which consider the potential launch sites of UAVs.

The two major parts of the framework, the mission planner representation and the defence planner representation, are shown in Fig.~\ref{fig:intent-framework}). These representations also allow the planner to generate labelled synthetic datasets representing various trajectory behaviours for training an intent recognition classifier. 
Different ways of mathematically defining the same informal objective of protecting the geo-fence are discussed.  
The framework is illustrated using a library of four intents in a security application. 
Numerical simulations are presented to illustrate various aspects of the intent inference framework. As part of this, an attention based Bi-directional LSTM network is presented for the intent classifier.
%\end{itemize}

In \ref{sec: systemModel}, the intent is defined mathematically and the concept of critical waypoint regions is illustrated using a library of four intents with  reference to a geo-fence in a $2D$ environment (though the framework is also applicable to 3D environments as shown later in the simulation study). %The motion models and path data can be produced in 3D and the framework as shown later in the simulation study, but for simplicity we present 2D examples in this paper. 
In \ref{sec: kinematic motion model}, the kinematic motion model and the methodology of motion pattern generation using  motion planning algorithms are described. In \ref{sec: objectives}, different ways of mathematically describing the objective are discussed. In \ref{sec:simulations}, simulations are presented to demonstrate various aspects of the intent inference framework.

% Please add the following required packages to your document preamble:
% \usepackage[table,xcdraw]{xcolor}
% Beamer presentation requires \usepackage{colortbl} instead of \usepackage[table,xcdraw]{xcolor}

\section{System Model}
\label{sec: systemModel}

Let $\mathcal{E}\subset \mathbb{R}^3$ be the environment which contains geo-fence $\mathcal{G}$ that is of interest. The environment also contains $\mathcal{B}$, which acts as an obstacle to the flight of aerial systems. So, $\mathcal{B},\mathcal{G} \subset \mathcal{E}.$ 

%$\mathcal{G}$ can be assumed to be a convex region.

\begin{figure*}[!t] % Use figure* to span both columns
    \centering
    \begin{minipage}{0.24\textwidth}
        \centering
        \includegraphics[width=\textwidth]{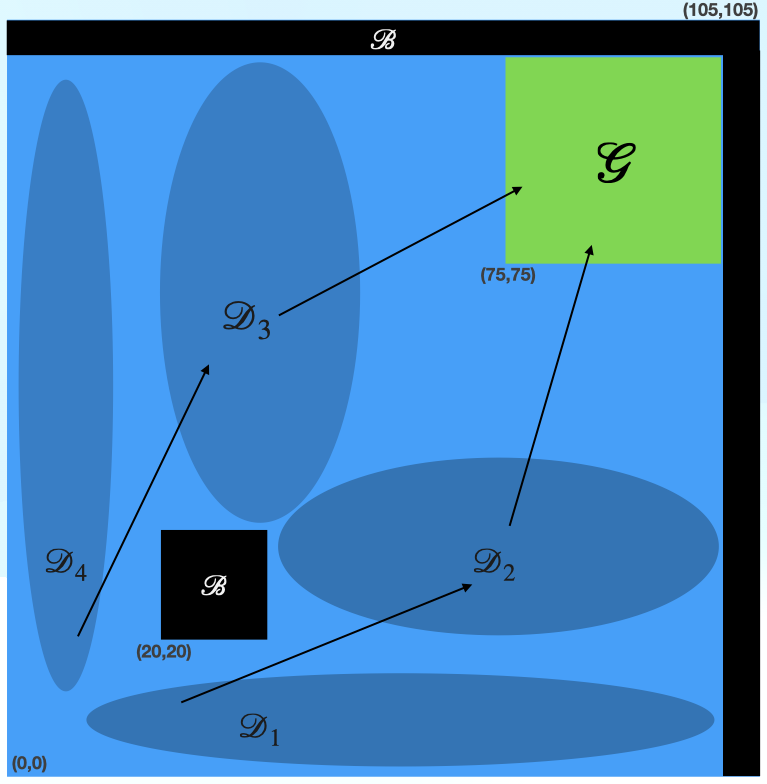}
        \caption*{(a) Direct Attack}
        \label{fig:da_ill}
    \end{minipage}\hfill
    \begin{minipage}{0.24\textwidth}
        \centering
        \includegraphics[width=\textwidth]{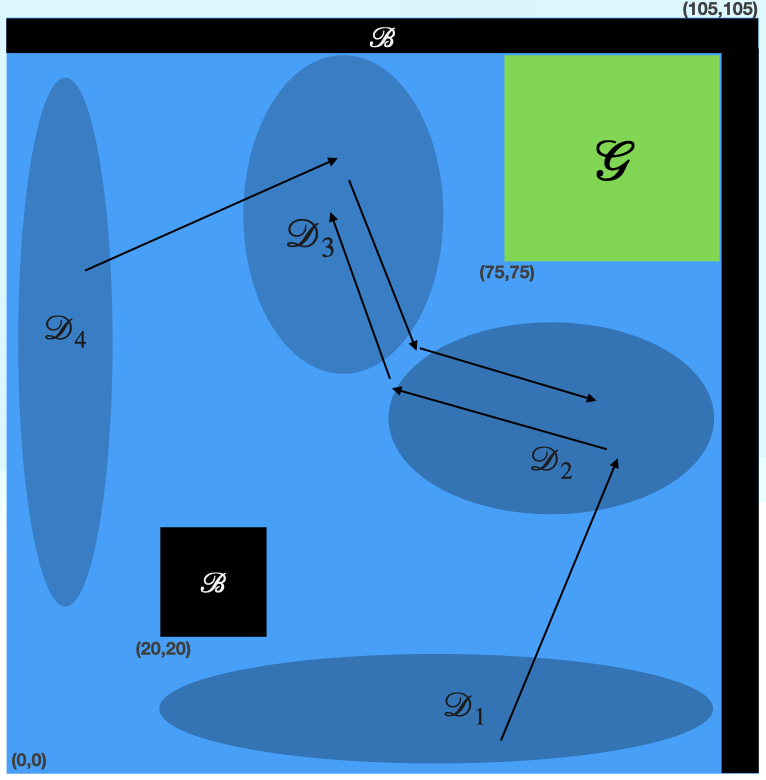}
        \caption*{(b) Surveillance}
        \label{fig:sur_ill}
    \end{minipage}\hfill
    \begin{minipage}{0.24\textwidth}
        \centering
        \includegraphics[width=\textwidth]{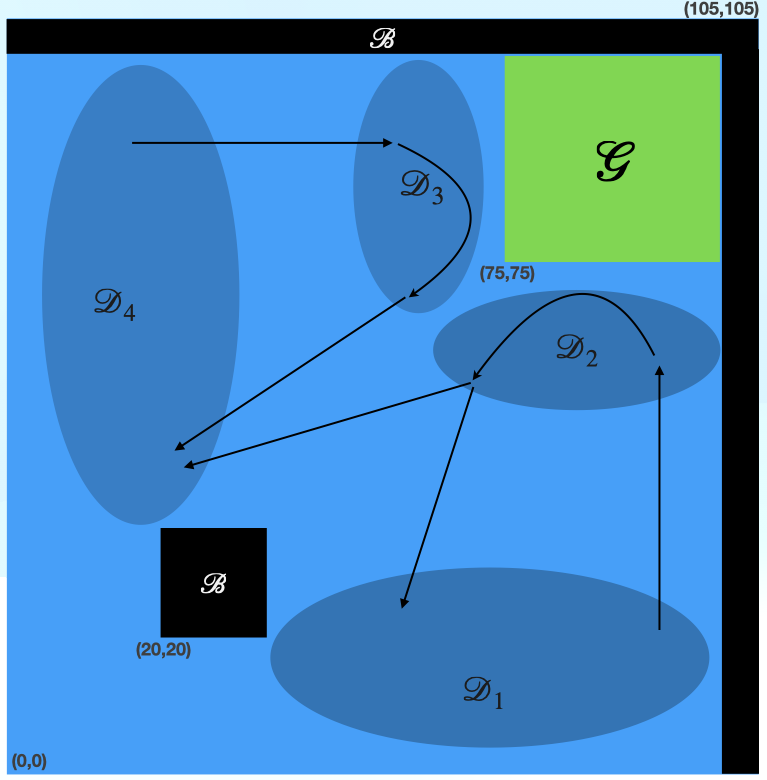}
        \caption*{(c) Criminal-Harmful}
        \label{fig:criminal_ill}
    \end{minipage}\hfill
    \begin{minipage}{0.24\textwidth}
        \centering
        \includegraphics[width=\textwidth]{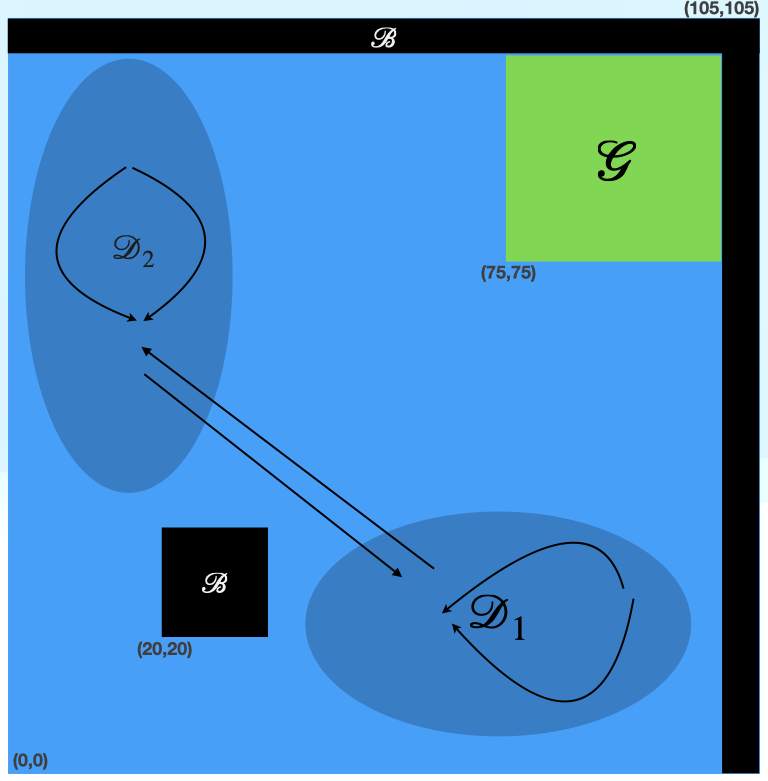}
        \caption*{(d) Harmless}
        \label{fig:harmless_ill}
    \end{minipage}
    \caption{A simple illustration of the critical waypoints patterns for various example intents. The oval shapes are arbitrary, chosen for the purpose of illustration. (a) Any system planning a direct attack enters the critical regions $\mathcal{D}_2$ or $\mathcal{D}_3$ with high probability. 
    (b) A UAV with surveillance intent might have a motion pattern moving from critical regions $\mathcal{D}_1$ to $\mathcal{D}_2$ to $\mathcal{D}_3$ or from $\mathcal{D}_4$ to $\mathcal{D}_3$ to $\mathcal{D}_2$, making an attempt of image acquisition from different perspectives.
    %$r_1\in\mathcal{D}_1,r_2\in\mathcal{D}_2,r_3\in \mathcal{D}_3$ or $r_1\in\mathcal{D}_4,r_2\in\mathcal{D}_3,r_3\in \mathcal{D}_2.$ 
    %
    (c) A criminal-harmful UAV has a motion pattern might suggest potentially harmful activity such as drop off or pick up, it gets close to the geo-fence temporarily and finally escapes. For example, it may go from $\mathcal{D}_1$ to $\mathcal{D}_2$ or from $\mathcal{D}_4$ to $\mathcal{D}_3$ for drop off, and then flee to $\mathcal{D}_4$ to $\mathcal{D}_1.$
    (d) A clueless-harmless UAV might have a motion pattern that stays in or moves between critical regions $\mathcal{D}_1 $ and $\mathcal{D}_2.$
    %such that $r_1,r_2,r_3\in \mathcal{D}_1 \text{ or } \mathcal{D}_2.$
    }
    \label{fig:intentlibrary_illust}
\end{figure*}

\subsection{Intent Models}
Some typical qualitative terms for intents that are commonly related to geographical-site security applications include \emph{head-on attack, payload drop, surveillance, criminal, harmless}, etc. Each intent can be associated with a particular set of motion patterns. Alternatively, each intent can be associated with a distribution of possible intermediate and final destination sequences. Although most planners might agree with the qualitative description of the above terms, they might associate the same terms with slightly different motion patterns as necessary for their particular situation. For example, surveillance motion patterns can differ based on the type of facility, such as a stadium, office building, etc.

To shift this discussion to a quantitative format, we mathematically define \textit{intent} in the following way. To efficiently describe the motion pattern we will use intermediate destinations as latent variables. 

%\textit{Intent:} An intent $I\in \mathcal{I}$ is a latent variable which is defined by $(\mathcal{E},\mathcal{G},N,\mathbb{P}_I).$ Here, $N$ is a positive integer, $\mathbb{P}_I: \mathcal{E}^N \to [0,1],$ and $\mathbb{P}_I(\mathbf{r})$ is the joint probability of the sequence $\mathbf{r}\coloneqq \{r_1,r_2,\ldots,r_N\}.$ 

%For convenience, let $\mathbf{r}\coloneqq \{r_n\}_{n=1:N}.$
%described by the sequence $\{r_1,r_2,\ldots, r_n\}$ the transition law $p(r'|r,z)$

\begin{definition}
\textit{Intent} :-- An intent $I\in \mathcal{I}$ of aerial system $AS$ is characterized by the entities $(\mathcal{E}, \mathcal{G}, \mathbb{N},\mathbf{P}_I, {\Phi}_I)$ which are defined as follows. 
\begin{itemize}
    \item $\mathcal{E}\subset \mathbb{R}^3$ is an environment.
    \item $\mathcal{G}\subset \mathcal{E}$ is a geo-fence inside the environment.
    \item $\mathbb{N}$ is an index set.
    \item $\mathbf{P}_I$ is called the critical waypoint pattern (CWPP) of intent $I.$ It is a stochastic process $\{\mathbf{r}_n,n\in\mathbb{N}\}$ where $\mathbf{r}_n$ is a random variable defined on the probability space $(\mathcal{E},\mathcal{E}_{\sigma},\mathbb{P}).$ Here, $\mathcal{E}_{\sigma}$ is a Borel $\sigma-$algebra of the subsets of $\mathcal{E}$ and $\mathbb{P}$ is a probability function.
    \item $\Phi_I$ is called the motion process of $I.$ It is a dynamical system defined on the kinematic state space of $AS,$ and is described by a discrete time difference equation $\mathbf{s}(k+1) = \Phi_I(\mathbf{s}(k)).$ The state space is $\mathcal{E}\times CS$ where $CS$ is the configuration space of $AS$ specifying the range of speed, acceleration, turn rate for each degree of freedom. We have ${s}(k) \coloneqq (\hat{\mathbf{r}}(k),\kappa(k)), k\geq 0$ with $\hat{\mathbf{r}}(k) \in \mathcal{E}, \kappa(k) \in CS.$ 
\end{itemize}
\end{definition}
Here, $\mathcal{I}$ is the library of intents defined with reference to environment $\mathcal{E}$ and geo-fence $\mathcal{G}.$ The CWPP describes the pattern of critical waypoints (CWPs) which are important waypoints essential for carrying out the intent. The motion process describes the kinematic motion of $AS$ in the environment. 
Intents are well-defined if the CWPP and the motion process are related in the following way. There exist $k_n>0$ for each $n\in \mathbb{N}$ such that $\mathbb{P}(\Vert \mathbf{r}_n - \hat{\mathbf{r}}(k_n) \Vert_2 < \epsilon)>1-\delta_m,$ where $\epsilon,\delta_m >0$ are arbitrarily small, predefined constants. This means that if critical waypoint sequence is specified, the motion process can generate a trajectory that takes $AS$ into the $\epsilon-$neighbourhood of those critical waypoints with high probability.

To specify stochastic process CWPP, the distributions of all random variables $\{\mathbf{r}_n\},n\in\mathbb{N}$ must be specified. A simple way of specifying these distributions is by defining critical waypoint regions or critical regions where critical waypoints are found almost surely.
%These regions are defined as follows.

\begin{definition}
    \textit{Critical Region Ordered Set (CROS)} :-- Given a set of regions $DR = \{\mathcal{D}_1,\mathcal{D}_2,\ldots, \mathcal{D}_m \}\subset \mathcal{E} $ for some integer $m$ such that $\mathcal{D}_i \cap \mathcal{D}_j=\phi (\mathsf{null set})$ for $i\neq j.$ A CROS is defined as $\mathcal{D} = \{\mathcal{D}^c_n\}, n \in \{1,2,\ldots,\vert \mathbb{N} \vert\}, \mathcal{D}^c_n\in DR,$ such that $\mathbb{P}(\mathbf{r}_n\in \mathcal{D}^c_n)>1-\delta_w$ for $n\in \{1,2,\ldots, \vert \mathbb{N} \vert\},$ $\vert \mathbb{N} \vert\leq m$ and an arbitrarily small $\delta_w>0.$ The individual elements of set $\mathcal{D}$ are called critical regions or critical waypoint regions.
\end{definition}
%
%\begin{definition}
%    \textit{Critical Region Ordered Set (CROS)} :-- A CROS is defined as $\mathcal{D}\subset \mathcal{E},$ $\mathcal{D} = \{\mathcal{D}_1,\mathcal{D}_2,\ldots, \mathcal{D}_{\vert \mathbb{N} \vert} \}$ with $\mathcal{D}_i \cap \mathcal{D}_j=\phi$ for $i\neq j$ such that $\mathbb{P}(\mathbf{r}_n\in \mathcal{D}_n)>1-\delta_w$ for $n\in \{1,2,\ldots, \vert \mathbb{N} \vert\}$ and an arbitrarily small $\delta_w>0.$ The individual elements of set $\mathcal{D}$ are called critical regions or critical waypoint regions.
%\end{definition}
%

For a given intent there might exist multiple CROS. Also, different intents can have similar CROS but may differ in the kinematic state sequences dictated by their motion processes.
%This definition has the advantage that the sequence $\mathbf{r}$ can be specified by a human pilot or derived from a path planning algorithm. 

%Assuming that the next destination depends on the The joint probability distribution $\mathbb{P}_I(\mathbf{r})$ can be specified by $$
%Motion of the agent under an intent is described by 

%Here, all points in the sequence $\mathbf{r}$ need not always be way points because the value of latent variable can change before reaching specified destination. This might change the trajectory of the agent towards another destination. 

To illustrate the intent model, we present an example intent library with $4$ intents defined with reference to a $2D$ region of interest. 
These are commonly employed intents in a security application.  But, a planner can make up their own library of intents and  intent models by using the above definition.

\subsubsection*{Example 1: Bounded 2D Environment}
Let the environment $\mathcal{E} = \{(x,y) \in [0,105]^2\},$ the geo-fence $\mathcal{G}= \{(x,y)\in [75,100]^2\}$ and obstacles $ \mathcal{B} = \{(x,y) \in [20,30]^2 \cup [0,105]\times [100,105] \cup [100,105]\times [0,105]\}.$ 

%\begin{enumerate}
    %
    1) \emph{Direct Attack or Payload drop}: Suppose the goal of the $AS$ agent is to launch a head-on attack or drop off payload at a location inside the geo-fence. Fig.~\ref{fig:intentlibrary_illust}(a) illustrates a motion pattern that suited this intent. Here, in order to attack $\mathcal{G}$ while avoiding the obstacles, an $AS$ almost surely enters regions $\mathcal{D}_2$ or $ \mathcal{D}_3.$ So we can describe the CWPP by defining $\mathbb{N}=\{1,2,3\} $ and 
    $\mathbf{r}_1 \in \mathcal{D}_1 \cup \mathcal{D}_4$, $\mathbf{r}_2 \in \mathcal{D}_2 \cup \mathcal{D}_3$, and $\mathbf{r}_3 \in \mathcal{G}$. 
    So, critical region ordered sets can be defined as $\{\mathcal{D}_1,\mathcal{D}_2,\mathcal{G}\}$ and $\{\mathcal{D}_4,\mathcal{D}_3,\mathcal{G}\}$.
    Alternatively, the planner might also use $\mathbb{N}=\{1,2,3,4\}$ with CROS' = $\{\mathcal{D}_1,\mathcal{D}_2,\mathcal{D}_3,\mathcal{G}\}$ and  $\{\mathcal{D}_4,\mathcal{D}_3, \mathcal{D}_2, \mathcal{G}\}$ depending on the planner's requirement.
    %So we can describe the CWPP by defining $\mathbb{N}=\{1,2,3\} $ and $\mathbf{r}_1,\mathbf{r}_2\in \mathcal{D}_1\cup \mathcal{D}_2\cup \mathcal{D}_3$ and $\mathbf{r}_3\in\mathcal{G}$. 
    %
    %So, critical region ordered sets can be defined as $\{\mathcal{D}_1,\mathcal{D}_1,\mathcal{G}\}$ or $\{\mathcal{D}_1,\mathcal{D}_2,\mathcal{G}\}$ or $\{\mathcal{D}_2,\mathcal{D}_1,\mathcal{G}\}$ or $\{\mathcal{D}_3,\mathcal{D}_1,\mathcal{G}\}$ or $\{\mathcal{D}_3,\mathcal{D}_2,\mathcal{G}\},$ depending on the planner's requirement.
    %Set of destinations are on or inside the geo-fence. 
    %
    %\begin{figure}[!h]
    %\centering
    %\includegraphics[width=0.8\linewidth]{figures/DirectAttackIntent.png}
    %\caption{Example 1: Direct Attack Intent in 2D environment. Any system planning a direct attack must inevitably enter the critical regions $\mathcal{D}_1$ or $\mathcal{D}_2.$}
    %\label{fig:directattack2d}
    %\end{figure}
    %
    %\item  Payload drop: Drop off payload on a target inside geo-fence.
    %
    
    2) \emph{Surveillance}: Suppose the mission of the $AS$ is to carry out surveillance of the target from a close-enough distance and gather footage from different perspectives. The critical waypoints would be along the geo-fence  as shown in Fig.~\ref{fig:intentlibrary_illust}(b). The motion pattern is such that the critical waypoint sequence is $\mathbf{r}_1\in\mathcal{D}_1,\mathbf{r}_2\in\mathcal{D}_2,\mathbf{r}_3\in \mathcal{D}_3$ or $\mathbf{r}_1\in\mathcal{D}_4,\mathbf{r}_2\in\mathcal{D}_3,\mathbf{r}_3\in \mathcal{D}_2$. So, CROS' = $\{\mathcal{D}_1,\mathcal{D}_2,\mathcal{D}_3\}$ and 
    $\{\mathcal{D}_4,\mathcal{D}_3, \mathcal{D}_2\}$.

    %
    %\begin{figure}[h]
    %    \centering
    %    \includegraphics[width=0.8\linewidth]{figures/Surveillance2D.png}
    %    \caption{Example 1: Surveillance Intent in 2D environment. The motion pattern is such that $r_1\in\mathcal{D}_1,r_2\in\mathcal{D}_2,r_3\in \mathcal{D}_3$ or $r_1\in\mathcal{D}_4,r_2\in\mathcal{D}_3,r_3\in \mathcal{D}_2.$}
    %    \label{fig:surveillance2d}
    %\end{figure}
    %
    
    3) \emph{Criminal-harmful}: The mission of the $AS$ is to drop off or pick up packages containing weapons or other potentially harmful materials on the edge of the geo-fence and then escape as illustrated in Fig.~\ref{fig:intentlibrary_illust}(c). The motion pattern is such that $\mathbf{r}_1\in \mathcal{D}_1,$ $\mathbf{r}_2\in \mathcal{D}_2,$ $\mathbf{r}_3\in \mathcal{D}_1$. 
    %For this mission $\mathbf{r}_2,\mathbf{r}_3$ are near the edge of $\mathcal{G}.$ 
    CROS' = $\{\mathcal{D}_1,\mathcal{D}_2,\mathcal{D}_1\}$, $\{\mathcal{D}_4,\mathcal{D}_3,\mathcal{D}_4\}$,  $\{\mathcal{D}_1,\mathcal{D}_2,\mathcal{D}_4\}$,  $\{\mathcal{D}_4,\mathcal{D}_3,\mathcal{D}_1\}$.  %
    %\begin{figure}[h]
    %    \centering
    %    \includegraphics[width=0.8\linewidth]{figures/Criminal-harmful-2D.png}
    %    \caption{Example 1: Criminal-harmful intent in 2D environment. The motion pattern is such that $r_1\in \mathcal{D}_1,$ $r_2 r_3\in \mathcal{D}_2$ or  $r_1\in \mathcal{D}_4,$ $r_2,r_3\in \mathcal{D}_3.$ $r_2, r_3$ is on the edge of $\mathcal{G}.$}
    %    \label{fig:criminal-harmful2d}
    %\end{figure}
    %\item Criminal-harmfull: The UAV might be conducting unlawful but not necessarily harmful activity such as rash movements transgressing the geo-fence temporarily.  
    %
    
    4) \emph{Harmless}: Harmless with no particular mission with respect to the geo-fence. The motion pattern is  shown in Fig.~\ref{fig:intentlibrary_illust}(d).
    %\begin{figure}[!h]
    %    \centering
    %    \includegraphics[width=0.8\linewidth]{figures/Clueless-harmless-2D.png}
    %    \caption{Example 1: Clueless-harmless intent in 2D environment. The motion pattern is such that $r_1,r_2,r_3\in \mathcal{D}_1 \text{ or } \mathcal{D}_2.$}
    %    \label{fig:clueless-harmless2d}
    %\end{figure}
    %
%\end{enumerate}

In the above discussion, we defined intent as a probabilistic motion pattern by means of a sequence of intermediate destinations. It is impractical to specify all the way points on every possible trajectory under an intent. In practice, the intermediate destinations can specified by simply specifying important regions where trajectory changes or maneuvering happens. To dictate the trajectory and kinematic parameters between the specified intermediate destinations, motion planning algorithms can be utilized. In the next section we describe a dynamic model which describes the evolution of kinematic states of the $AS.$ 

\section{Kinematic motion model}
\label{sec: kinematic motion model}
The kinematic state vector $s$ of the agent at time $t$ consists of the position, velocity and acceleration variables.
{\small{
\begin{align}
\label{eqn: state vector}
%s(t) = [x(t),v_x(t),a_x(t),y(t),v_y(t),a_y(t)]^T
s(t) = [x(t),v_x(t),a_x(t),y(t),v_y(t),a_y(t),z(t) ,v_z(t),a_z(t)]^T 
\end{align}
}}
As the defence planner has no access to the agent's exact motion model, it assumes that the agent can be in a finite set of flight motion models or modes, whose dynamic models are described below. The following modes correspond to fixed wing UAVs, but, other modes can be defined as required by planner. The dynamical models of these modes are utilized by the planner for feature extraction. The trajectory information is passed into an IMM filter along with the dynamical models of the modes. This filter is  designed to tracking highly maneuverable objects using multiple motion models. 
%The motion planning algorithm has to select the optimal mode and its parameters to traverse between critical waypoints.
%
To illustrate the structure of flight modes we discuss the modes of a fixed wing UAV.
\subsection*{Flight Modes}
The basic flight modes include the constant velocity (CV) and constant acceleration (CA) modes. The horizontal motion with minimum vertical maneuvers can be represented by the Horizontal Coordinated Turn (HCT) model. If there are maneuvers in 3D space, the 3D Coordinated Turn (3DCT) model can be used instead. The set of modes is $\mathcal{M}=\{CV,CA,HCT,3DCT\}.$
We describe the dynamic model for each mode in the following.

{\small{
\begin{align}
\label{eqn: dynamic model}    
    s(t_k) &= F_{m} s(t_{k-1}) + w(t_k),\\
    &m \in \{\text{CV, CA}\}.\nonumber\\
    \begin{bmatrix}
\label{eqn: dynamic model2}    
    s(t_k) \\ \omega_k     
    \end{bmatrix} &= F_{m} \begin{bmatrix}
        s(t_{k-1}) \\ \omega_{k-1}
    \end{bmatrix}  + w'(t_k),\\
    &m \in \{\text{HCT, 3DCT}\}.\nonumber
\end{align}
\begin{align}
A_{CV} = \begin{bmatrix}
1 & T & 0\\
0 & 1 & 0\\
0 & 0 & 0
\end{bmatrix}, 
\text{  }
F_{CV} = \begin{bmatrix} 
A_{CV} & 0 & 0 \\ 
0 & A_{CV} & 0 \\ 
0 & 0 & A_{CV} 
\end{bmatrix} \\
%\end{equation}
%
%\begin{equation}
A_{CA} = \begin{bmatrix}
1 & T & \frac{T^2}{2}\\
0 & 1 & T\\
0 & 0 & 1
\end{bmatrix}, 
\text{  }
F_{CA} = \begin{bmatrix} 
A_{CA} & 0 & 0 \\ 
0 & A_{CA} & 0 \\ 
0 & 0 & A_{CA} 
\end{bmatrix}
\end{align}
\begin{align}
\begin{aligned}
A_{HCT}(\omega) &= \begin{bmatrix}
1 & \frac{sin(\omega t)}{\omega} & 0 & 0 & \frac{cos(\omega T)-1}{\omega} & 0\\
0 & cos(\omega t) & 0 & 0 & -sin(\omega t) & 0\\
0 & -\omega sin(\omega t) & 0 & 0 & -\omega cos(\omega t) & 0\\
0 & \frac{1-cos(\omega T)}{\omega} & 0 & 1 & \frac{cos(\omega T)}{\omega} & 0\\
0 & sin(\omega T) & 0 & 0 & cos(\omega T) & 0\\
0 & \omega cos(\omega T) & 0 & 0 & -sin(\omega t) & 0 
\end{bmatrix} \\
 F_{HCT}(\omega) &= \begin{bmatrix} 
A_{HCT}(\omega) & 0 & 0 \\ 
0 & A_{CV} & 0 \\ 
0 & 0 & 1 
\end{bmatrix}
\end{aligned}
\end{align}
\begin{align}
\begin{aligned}
A_{3DCT}(\Omega) &= 
\begin{bmatrix}
1 & \frac{sin(\Omega T)}{\Omega} & \frac{1-cos(\Omega T)}{\Omega^2}\\
0 & cos(\Omega T) &\frac{sin(\Omega T)}{\Omega}\\
0 & -\Omega sin(\Omega T) & cos(\Omega T)
\end{bmatrix} \\
%\text{  }
%\end{equation}
%
%\begin{equation}
F_{3DCT}(\Omega) &= \begin{bmatrix} 
A_{3DCT}(\Omega) & 0 & 0 & 0\\ 
0 & A_{3DCT}(\Omega) & 0 & 0\\ 
0 & 0 & A_{3DCT}(\Omega) & 0\\
0 & 0 & 0 & 1
\end{bmatrix}
\end{aligned}
\end{align}

}}
%
%
%\subsection{Multi-Mode Latent Destination Model}
%
%Given the trajectory points generated used for generation of state sequence as a function of the destination. 
The state sequence depends on the sequence of modes which in turn depend on the destination. Given state $s(t)$ and current mode $m(t)$ the next state is dictated by equations \eqref{eqn: dynamic model} and \eqref{eqn: dynamic model2}.
The sequence of modes themselves depend on the source and destination, and are decided by a motion planning algorithm. 
%The destination is not necessarily a way-point but serves to guide the trajectories. 
%The motion planning algorithms are described below.

\subsection*{Motion planning algorithm}
Let the history of modes and states until time $t,$ be $\mathcal{H}_t\coloneqq\{s(t'),m(t')\}_{t'=1:t}.$ Given the history, a motion planning algorithm $\phi : \mathcal{H}_t \mapsto \mathcal{S}$ outputs the next state (determined by the mode chosen). Two approaches to motion planning algorithms - 1) \textit {a priori} path finding  followed by trajectory tracking (path following), 2) hybrid \textit{dynamic} path finding and following algorithms.
%
%can be used here considering two types of scenarios - 1) perfect localization and target (destination) tracking, 2) imperfect localization, imperfect tracking capabilities and dynamic environment. 
For the first case a graph search based path finding algorithm can be used (see \cite{deFilippis:JIRS12:3Dpathplanning,Koenig:JAI04:LPA*}). For the second case a Markov decision process (MDP) based motion planning algorithms can be used. For the other scenarios involving uncertainty, a partially observable MDP based motion planning can be considered (see \cite{Ragi:AES13:UAV_POMDP,meuleau2010pomdp,miller2009pomdp}). As the state space and action space are quite large for a 3D environment, computation of optimal policy by value iteration is not practical. Instead, a near optimal policy can be learnt by a Deep Q-network as in \cite{Tu2023:DQNpathplanning}. 

Again, the planner can choose an algorithm as per their requirements. In our simulation experiments we choose RRT* algorithm for path finding, without any loss of generality.  

%Fig.~\ref{fig:mdp_planning} shows the modules involved in this approach.
%
%\begin{figure}
%    \centering
%    \includegraphics[width=\columnwidth]{figures/MDP based planning.png}
%    \caption{Modules involved in MDP based sequential motion planning.  }
%    \label{fig:mdp_planning}
%\end{figure}
%
%\subsubsection{Graph search based motion planning algorithm}
%Define graph $G=(\mathcal{S},E)$ with the set of edges $E \coloneqq \{ (u,v)\in \mathcal{S}^2 \vert \}.$ Let $r_D \coloneqq [x_D,y_D,z_D]$ be the destination location.

\section{Representation of Objectives }
\label{sec: objectives}
The overall objective of protecting the geo-fence can be described mathematically in multiple ways, depending on the protection requirements and the assumptions about the  motion patterns for various intents. Different ways of formulating the intent inference problem are given part of the framework. The decision maker tasked with protecting the geo-fence can choose from these formulations. 
In the following we discuss different ways posing the problem as a cost minimization problem and also present loss functions that can be used for supervised training of intent classifiers.

%Let $\mathcal{I}$ be the set of possible intents. 
\subsubsection{Intent inference as multi-class classification problem}
The objective here is to minimize the expected mis-classification probability.
For supervised training of a neural network for multi-class classification a commonly used loss function is the categorical cross entropy function. For a single sample with true label $y=[\mathsf{1}_{y\in l}]_{l\in \mathcal{I}}$ and predicted label (probability vector) $\hat{y}\in [0,1]^M$ the loss is as follows.
\begin{align}
    \mathcal{L}_{CCE}(\Gamma, y, \hat{y}) = -\sum_{l\in\mathcal{I}} y_l \log{\hat{y}_l}
\end{align}
\subsubsection{Multi-class classification with class imbalance and asymmetric error costs}
In practice the costs associated with various errors are not equal. For example, mistaking a UAV with payload drop intent for criminal-harmless may be considered costlier than mistaking criminal-harmless as criminal-harmful or mistaking criminal-harmless as clueless. Further, there might be class imbalance owing to relative scarcity of data for certain intents. The objective here is to minimize total cost of mis-classification which is given as 
\begin{align}
\min\mathbf{E}[\sum_{l,l'\in \mathcal{I}} c(l,l')\mathsf{Pr}(y\in l,\hat{y}\in l')],
\end{align}
where $c(l,l)$ is the cost of mistaking intent $l$ for $l'.$
In such cases we can use the asymmetric focal loss (AFL) function introduced by Barnes and Henn in \cite{Barnes:SPIE23:AsymmetricFocalLoss} in supervised training. This is a variation of focal loss introduced Lin et al in their pioneering paper \cite{Lin:ICCV17:FocalLoss} which considered class imbalance in training data. The loss function is as follows.
\begin{align}
    \mathcal{L}_{AFL}(\Gamma,y,\hat{y}) = -\sum_{l\in\mathcal{I}} y_l (1-\hat{y}_l)^{\gamma_l} \log{\hat{y}_l},
\end{align}
where $\gamma_l$ is a hyper-parameter for intent $l\in\mathcal{I}.$ 
\subsubsection{Probabilistic protection requirement}
One specification for intrusion protection is to minimize the probability of intrusion into the geo-fence $\mathcal{G}$. This requirement can be described as follows. 
\begin{align}
    \min \sum_{l\in\mathcal{I}} c_l^{\textup{int}} \mathsf{Pr}\left( \Gamma_{l} \rightarrow \mathcal{G} \right)
\end{align}
Here, $c_l^{\textup{int}}$ is the cost of intrusion with intent $l,$ and $\mathsf{Pr}\left( \Gamma_{l} \rightarrow \mathcal{G} \right)$ is the probability of intrusion of a trajectory of intent $l$ into $\mathcal{G}.$ This problem can be posed as multi-label classification problem with each state sequence $\Gamma$ has two labels, one corresponding to the intent and the second corresponding to whether the trajectory intrudes $\mathcal{G}.$
%
%The loss function that needs to be minimized can be defined as follows.
%\begin{align}
%    \mathcal{L}(\Gamma,y,\hat{y}) = \mathcal{L}_{CCE}(\Gamma, y,\hat{y}) + \mathcal{L}_{BCE}(\Gamma,z,\hat{z})
%\end{align}
%Here, $\mathcal{L}_{BCE}$ is the binary cross entropy function, $z \in\{0,1\}$ is the true observation that $\Gamma$ intrudes $\mathcal{G}$ and $\hat{z}\in [0,1]$ is the output probability of $\Gamma$ intruding $\mathcal{G}.$  

%In the above function the first is for the classification of  

\subsubsection{Time-constrained protection requirement}
Another way to prevent potential intrusions is to include an early classification requirement in the loss function. This can be done in the following way. $\Gamma$ denotes a time series of UAV states at $t=1,2,\ldots,T_{max}$ time instants. Let $\tau \leq T_{max}$ be the decision time when the intent classification is made and $T_{\textup{int}}(\Gamma)\leq T_{max}$ is the time $\Gamma$ intrudes $\mathcal{G}.$ If it doesn't intersect $\mathcal{G},$ then $T_{\textup{int}}(\Gamma)\rightarrow \infty.$ Let $\Gamma_{\rightarrow \tau}$ denote the partial state sequence observed until the decision time $t=\tau.$ The objective is to minimize the expected loss over all possible decision times.
\begin{align}
\begin{aligned}
    &\min \mathbf{E}_{\tau}[\mathcal{L}_{\tau}(\Gamma_{\rightarrow \tau},y,\hat{y})] \\
    \mathcal{L}_{\tau}(\Gamma_{\rightarrow \tau},y,\hat{y}) &= \alpha \mathcal{L}_{CCE}(\Gamma, y, \hat{y}) + (1-\alpha)\frac{\tau}{T_{\textup{int}}(\Gamma)}
\end{aligned}
\end{align}
This function represents a trade off between intent classification accuracy and early classification to prevent intrusion.

\section{Simulation Experiments}
\begin{figure*}[!ht]
    \centering
    \includegraphics[width=1\linewidth]{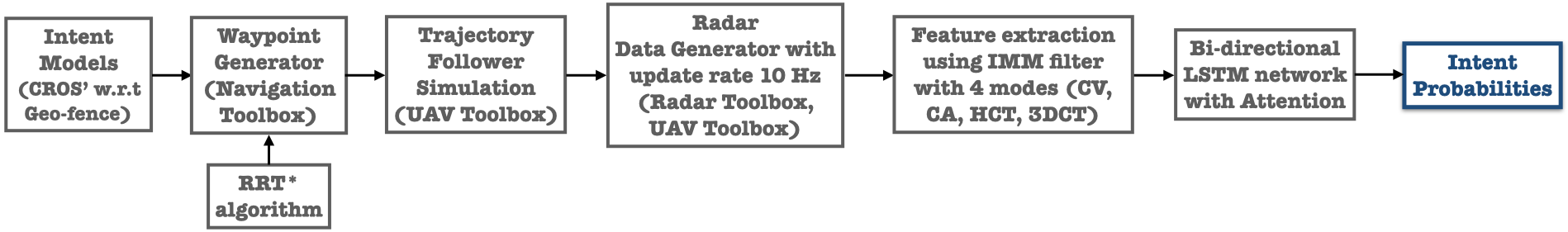}
    \caption{The various modules of the simulator. RRT* algorithm was used for path finding. The UAV trajectory and radar measurement simulation was implemented with MATLAB UAV and Radar toolbox.}
    \label{fig:simsetup}
\end{figure*}
\begin{figure}[htp]
    \centering
    \begin{tabular}{cc}
          \hspace{-0.25 in}

      \includegraphics[width=.5\linewidth]{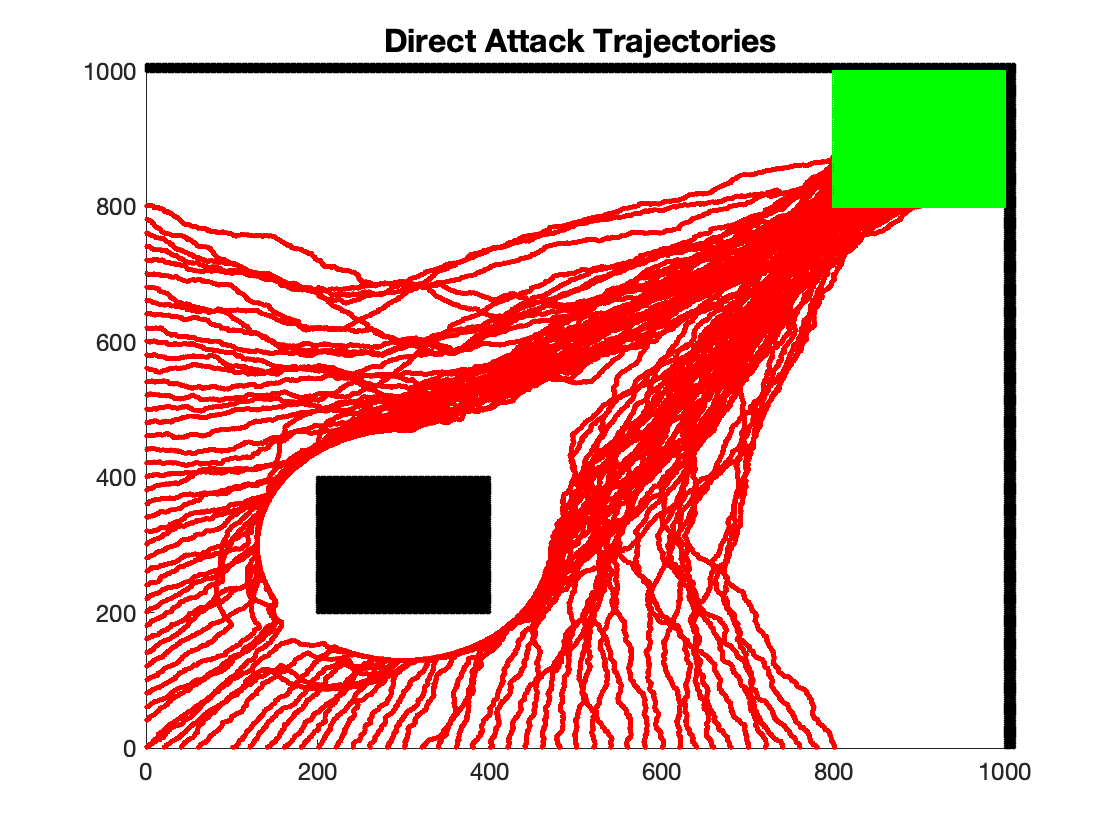} & 
      \hspace{-0.35 in}
      \includegraphics[width=.55\linewidth]{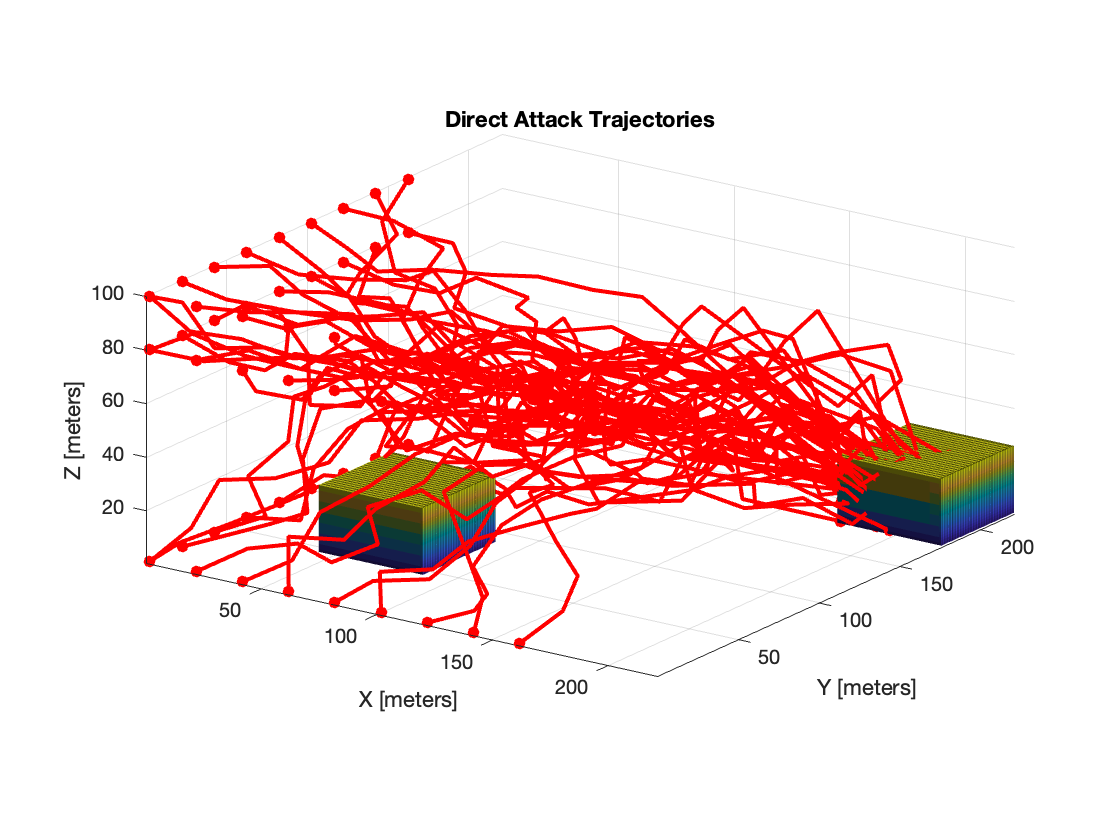}
      \\
      \hspace{-0.25 in}
      \includegraphics[width=.5\linewidth]{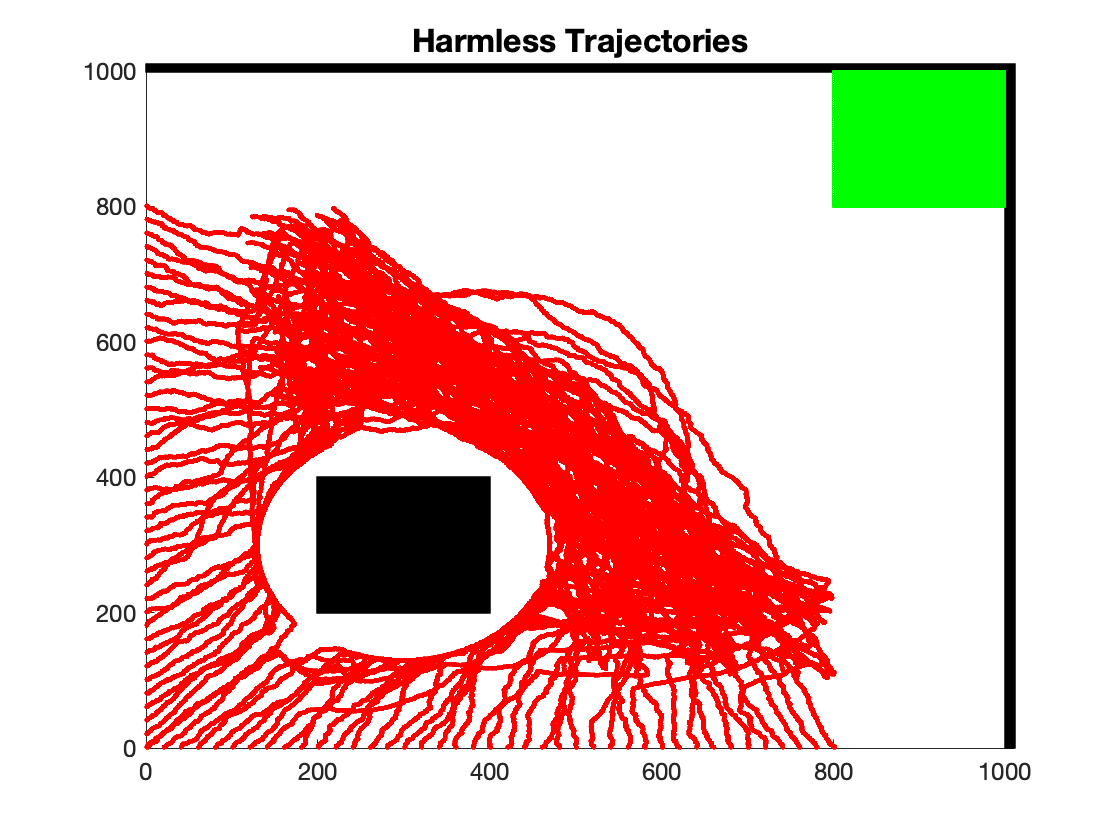} & 
            \hspace{-0.35 in}
      \includegraphics[width=.55\linewidth]{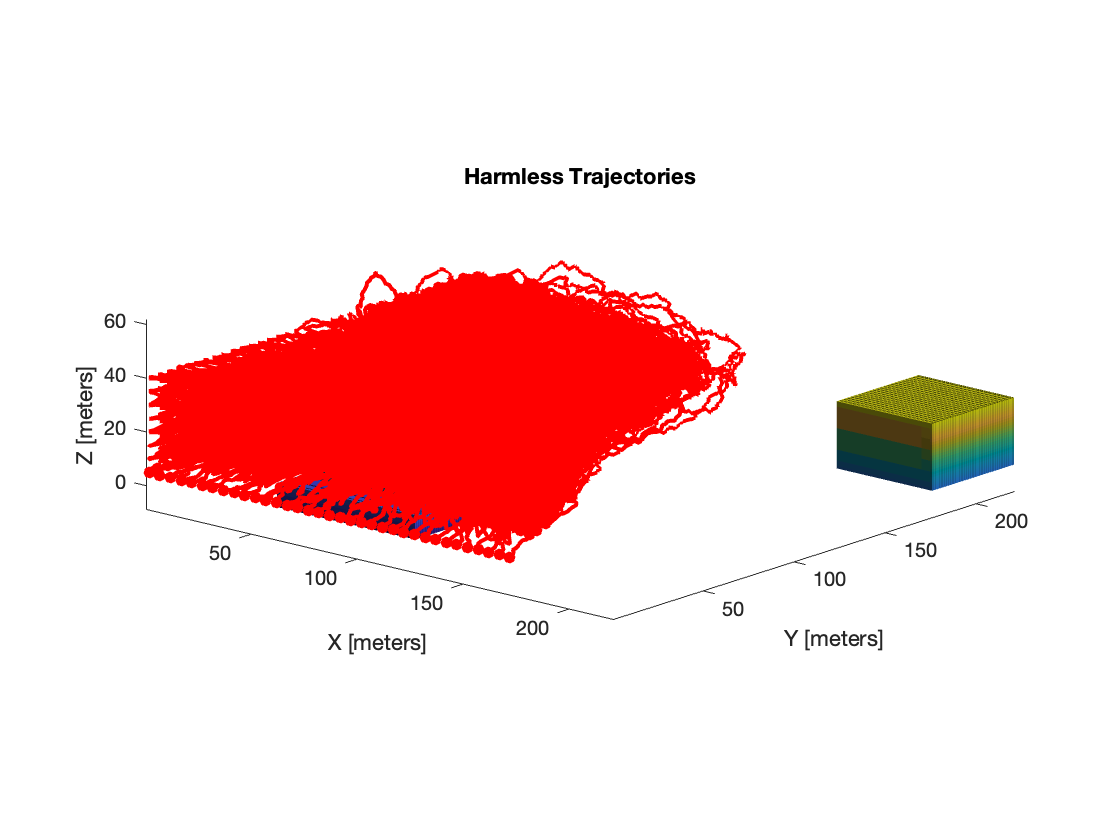}
      \\
      \hspace{-0.25 in}
    \includegraphics[width=.5\linewidth] {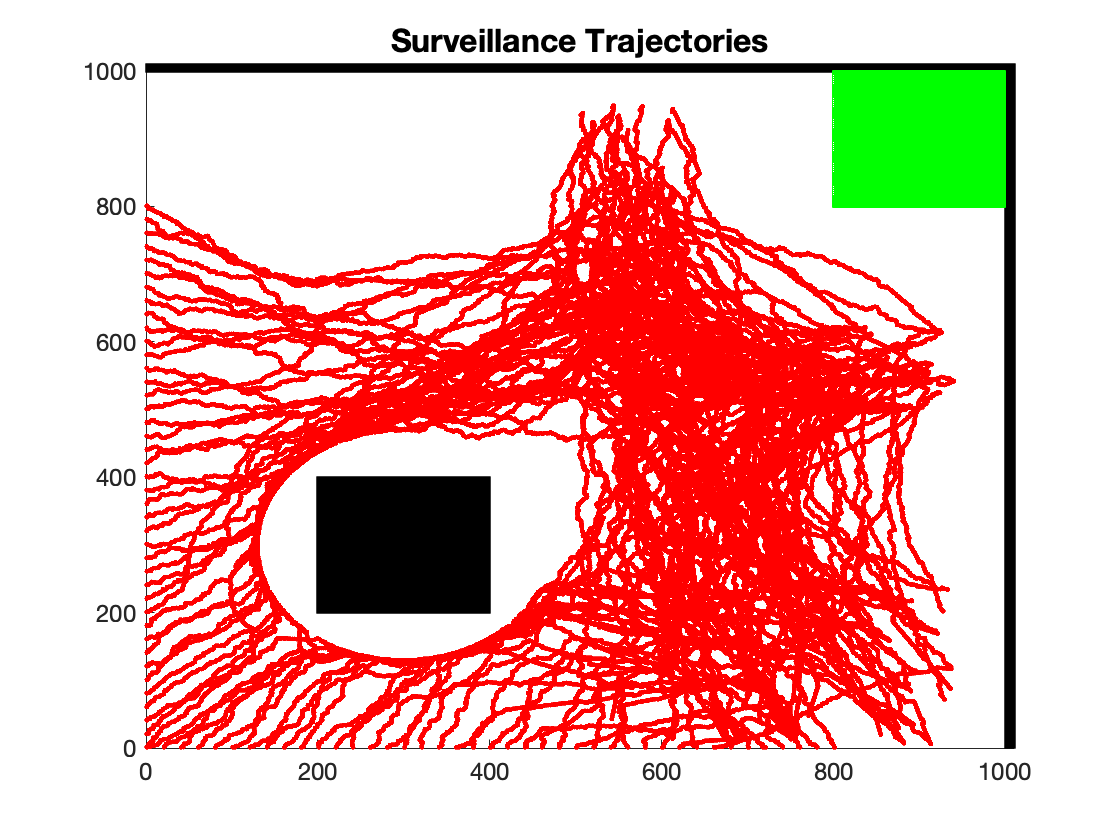} &       \hspace{-0.35 in}
    \includegraphics[width=.55\linewidth]{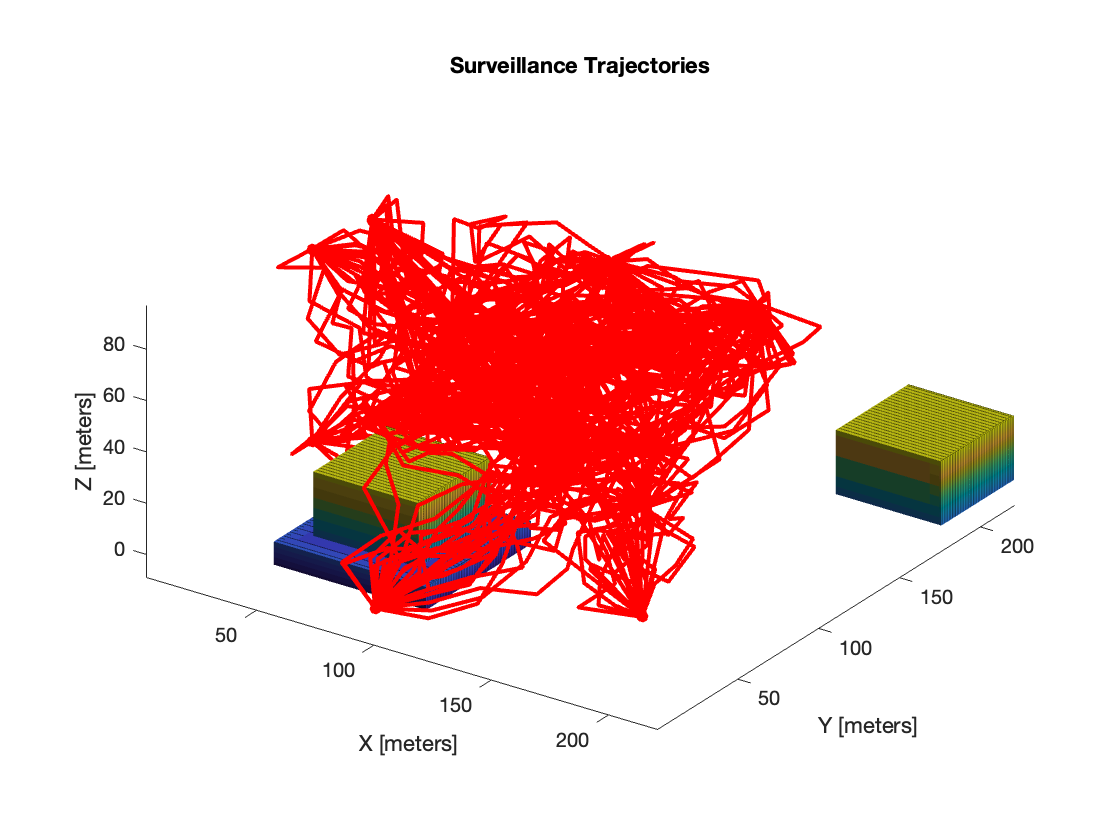}
        
    \end{tabular}
    \caption{Sample trajectories on 2D and 3D environments for various intents generated by the path finding algorithm RRT*.} %Direct attack and surveillance trajectories look similar in the initial parts, so the detector relies on other features such as velocity, acceleration and turn rate to infer the intent.}
    \label{fig: trajectories}
\end{figure}

\label{sec:simulations}
In the previous sections we presented a generic framework for intent inference and discussed the different parts of the framework through examples. We do not take up an elaborate simulation study that implements all aspects of the framework, and leave it for future work. We focus on intent inference as a multi-class classification problem (other possible objectives have been discussed in Section~\ref{sec: objectives}). The major modules involved in the simulation setup are summarized in Fig.~\ref{fig:simsetup}. 
\begin{figure}[!htp]
    \centering
    \begin{tabular}{cc}
    \includegraphics[width=0.45\columnwidth]{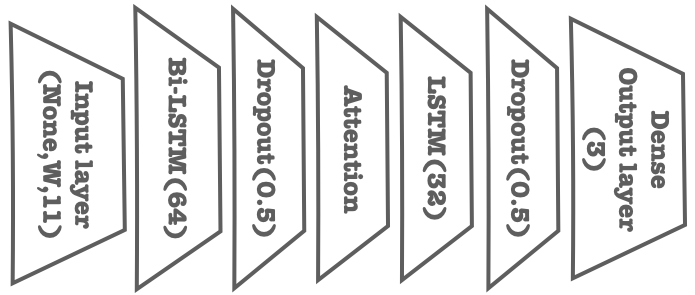} 
    &
   \includegraphics[width=0.48\columnwidth]{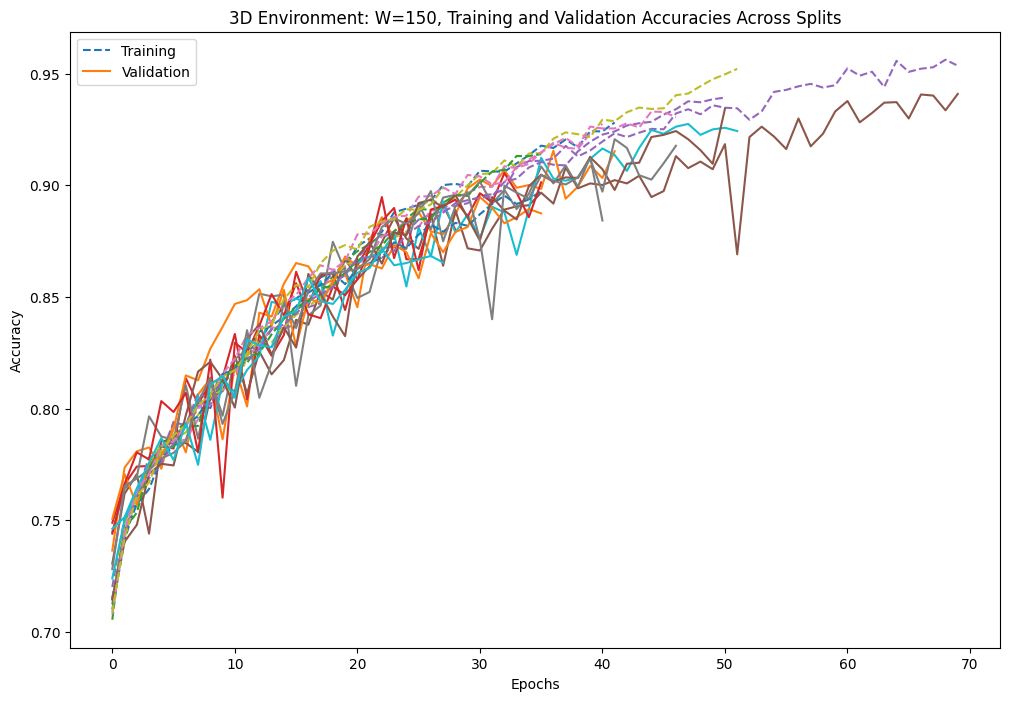}
    \end{tabular}
    \caption{[left] Architecture of the intent classifier. [right] Evolution of training and validation accuracy with number of epochs for various splits.}
    \label{fig:dnn}
\end{figure}
{{
\begin{table}[h]
\caption{Raw dataset size}
\label{tab:dataset_size}
\begin{center}
\begin{tabular}{|c|c|c|c|}
\hline
Intent & Direct Attack & Harmless & Surveillance \\ \hline
2D & 156 & 162 & 162 \\ \hline
3D  & 648 & 560 & 504 \\ \hline
\end{tabular}
\end{center}
\end{table}
}}
We considered 2D and 3D environments for generating the trajectory datasets. 
Labelled trajectory datasets were generated for different intents using by defining critical regions as described above and summarized in Fig.~\ref{fig:simsetup}. The number of trajectories in the dataset is given in Table~\ref{tab:dataset_size}. The difference in the number of samples for each intent is due to uniform sampling of initial points from differently sized critical waypoint regions. This class imbalance was normalized after feature extraction and the generation of the final dataset. These datasets were used to train an attention based Bi-LSTM network for intent recognition. The dimensions of the 2D and 3D environments are shown in Fig.~\ref{fig: trajectories}. 
\subsection{Classifier architecture}
The architecture utilizes an attention-based Bi-LSTM classifier. This approach takes advantage of Bi-LSTM's ability to capture dependencies in both forward and reverse directions of the time series data, providing a comprehensive understanding of the UAV's trajectory. The integration of the IMM filter with this classifier ensures that the dynamic state changes. For example, consider a UAV that starts circling over the restricted area; the IMM filter updates its motion models to reflect this maneuver, while the Bi-LSTM, with its attention mechanism, focuses on this significant behavioral change, interpreting it as potential surveillance activity. This allows for precise classification of the UAV’s intent, leveraging both past maneuvers and anticipating future movements. 

The neural network model is designed for sequential data with temporal dependencies, leveraging convolutional and LSTM layers wrapped in `TimeDistributed' to process each time step independently. It begins with a `TimeDistributed' 1D convolutional layer with 64 filters and a kernel size of 3, applied twice to capture local patterns within each time step. This is followed by a `TimeDistributed' dropout layer with a 0.5 rate to prevent overfitting, and a `TimeDistributed' max-pooling layer with a pool size of 2 to reduce dimensionality. The sequence is then flattened using a `TimeDistributed' flatten layer and fed into an LSTM layer with 20 units, which captures temporal dependencies across the sequence. The output of the LSTM is processed by a dense layer with 100 units and ReLU activation, followed by a final dense layer with 3 units and softmax activation to produce a probability distribution over three classes. The model is compiled with categorical crossentropy loss, the Adam optimizer, and accuracy as the performance metric. The classifier network is summarized in Fig.~\ref{fig:dnn}.

\subsection{Radar measurements and non-optimal paths}
\begin{figure}[htp]
    \centering
    \begin{tabular}{ccc}
      \hspace{-0.2 in}
      \includegraphics[width=.36\linewidth]{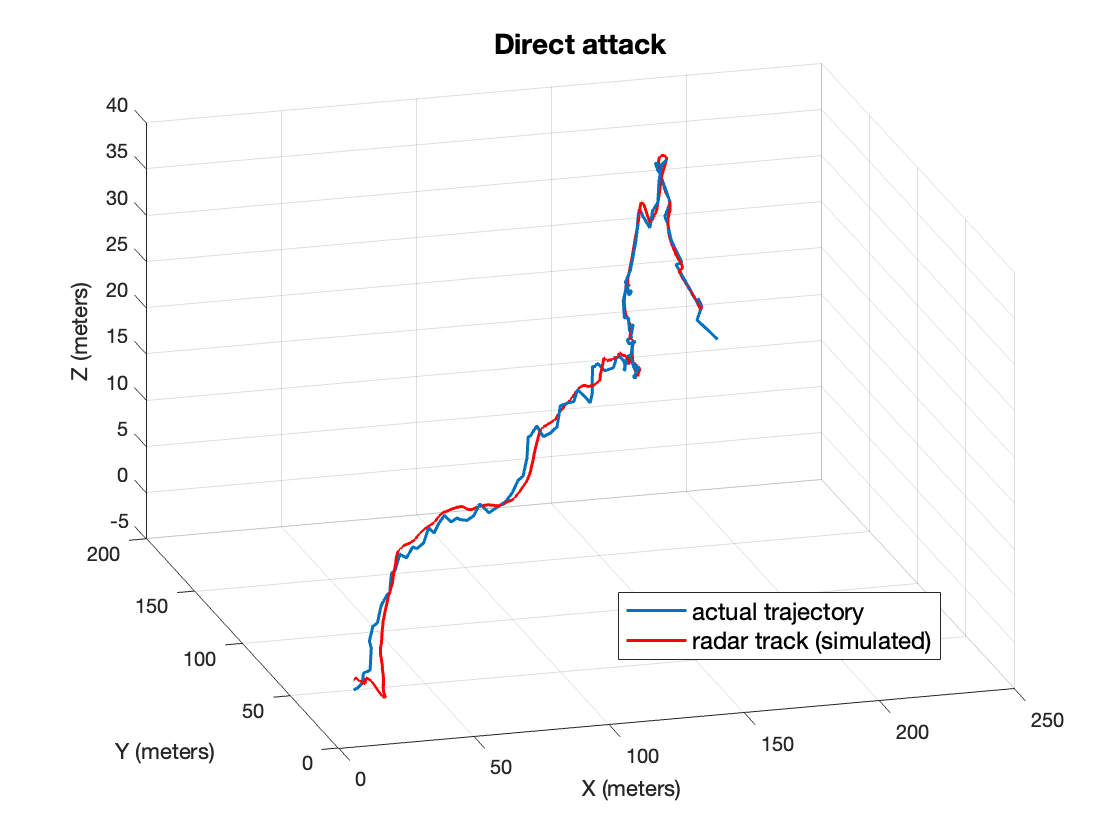} & 
      \hspace{-0.25 in}
      \includegraphics[width=.36\linewidth]{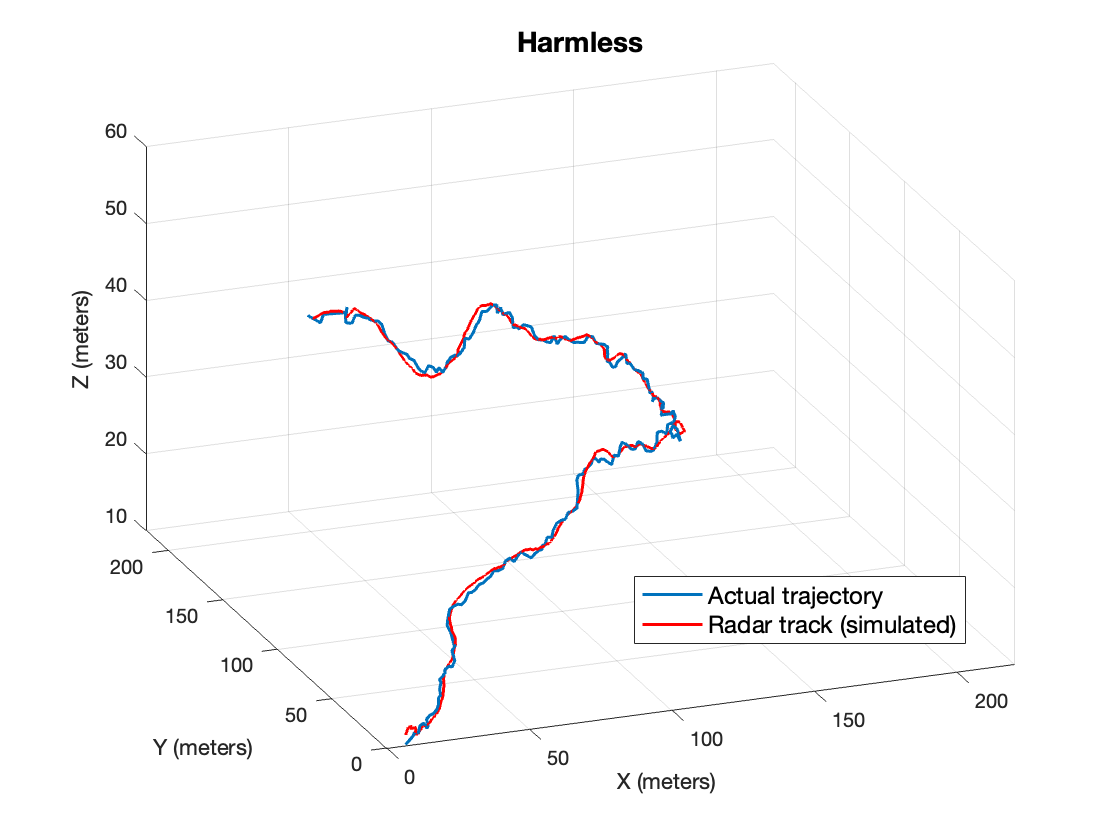} &
      \hspace{-0.3 in}
      \includegraphics[width=.36\linewidth]{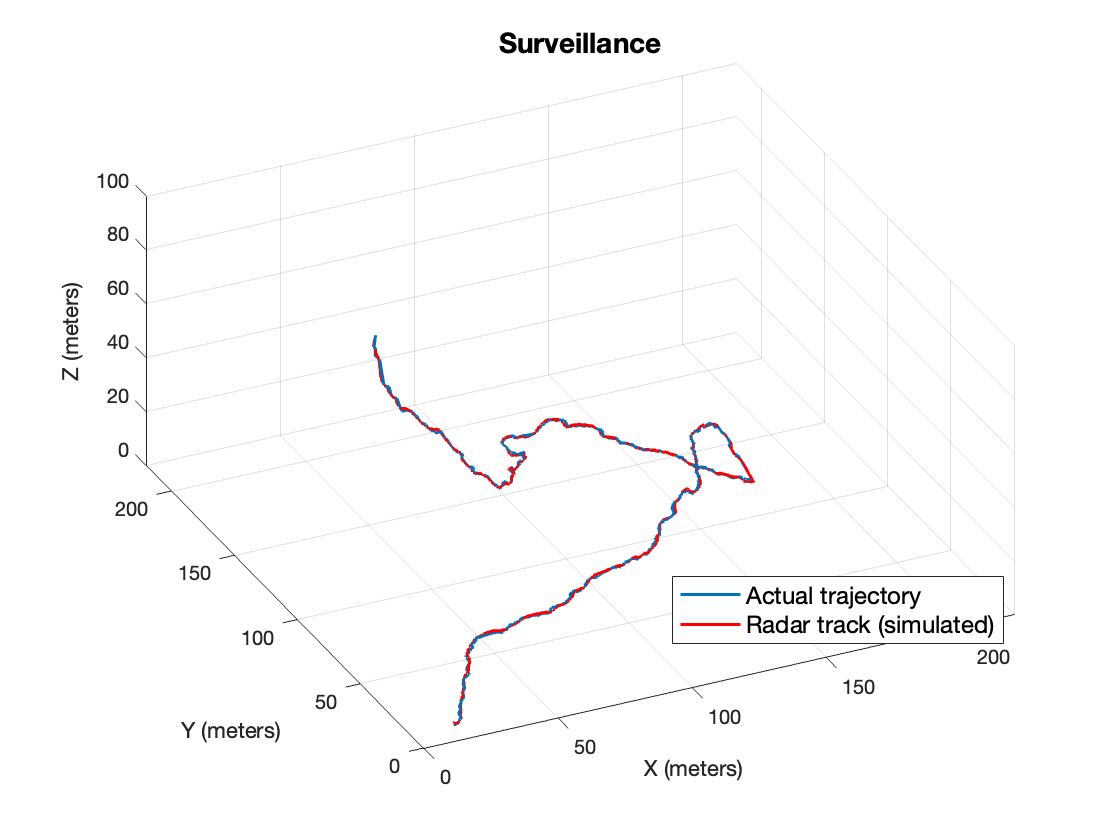}
        
    \end{tabular}
    \caption{Examples of trajectories along with the simulated radar tracks with a range error which falls within $2$ meters once the tracking filter is stabilized.} %Direct attack and surveillance trajectories look similar in the initial parts, so the detector relies on other features such as velocity, acceleration and turn rate to infer the intent.}
    \label{fig: tracks}
\end{figure}

In this experiment we use non-optimal path finding algorithms to generated the trajectories. Further, we assume that there is a radar inside the geo-fence which detects and tracks targets. Radar measurements are simulated using MATLAB UAV and radar toolboxes.  ``RadarDataGenerator'' parameters include ``FieldOfView'' = $[90,90]\degree$, ``CenterFrequency'' = $24.55$GHz, ``Bandwidth'' = $45$MHz, ``ReferenceRCS'' = $0\text{dBm}^2$, ``ReferenceRange'' = $2500$m, ``FalseAlarmRate'' = $10^{-6}$, ``ElevationResolution'' = $1$m, ``AzimuthResolution'' = $1$m, ``RangeResolution'' = $1$m.
Fig.~\ref{fig: tracks} shows example trajectories of different intents, their radar tracks. Simulated tracks showed a range error  around 10 meters in the initial parts of the trajectories; once the tracking filter stabilizes the error falls within $2$ meters. 
Tracking in done using Kalman filter and feature extraction is done using IMM filter with the flight mode models described earlier. Features (directional velocities, accelerations and turn rates) are extracted from radar measurements. These features are used to train the classifier. Three parametric intent scenarios are considered for simulation. The Direct Attack, Harmless and Surveillance have  speed ranges $15-25,$ $5-12,$ $5-15$ m/s, respectively.
%given in Table~\ref{tab: velocity ranges}.

%\begin{table}[]
%\caption{Velocity ranges ($m/s$) for various segments of the trajectories. A trajectory may be divided into segments by its critical way points. }
%\centering
%\begin{tabular}{|c|c|c|c|}
%\hline
%   & \begin{tabular}[c]{@{}c@{}}Segment 1\\ velocity ($m/s$)\end{tabular} & \begin{tabular}[c]{@{}c@{}}Segment 2\\ velocity ($m/s$)\end{tabular} & \begin{tabular}[c]{@{}c@{}}Segment 3\\ velocity ($m/s$)\end{tabular} \\ \hline\hline
%Direct Attack & $5-15$  & $15-25$   & $15-25$       \\ \hline
%Surveillance & $5-12$   & $5-12$    & $5-12$         \\ \hline
%Harmless    & $5-15$   & $5-15$      & $5-15$        \\ \hline
%\end{tabular}
%\label{tab: velocity ranges}
%\end{table}


%\begin{table}[]
%\caption{Sizes of the datasets}
%\begin{tabular}{|c|c|c|c|l}
%\cline{1-4}
% Intent & Trajectories & Feature vectors &   &  \\ \cline{1-4}
% Direct Attack &  &  &  &  \\ \cline{1-4}
% Harmless &  &  &  &  \\ \cline{1-4}
% Surveillance &  &  &  &  \\ \cline{1-4}
%\end{tabular}
%\end{table}
%
%
%\begin{table}[]
%\caption{Raw dataset size}
%\label{tab:dataset_size}
%\begin{center}
%\begin{tabular}{|c|cc|cc|}
%\hline
%Intent & \multicolumn{2}{c|}{Trajectories}                       & \multicolumn{2}{c|}{Feature vectors}        \\ \hline
%       & \multicolumn{1}{c|}{3D} & 2D             & \multicolumn{1}{c|}{3D} & 2D \\ \hline
%Direct Attack     & \multicolumn{1}{c|}{648}  &   156               & \multicolumn{1}{c|}{}           &           \\ \hline
%Harmless     & \multicolumn{1}{c|}{560}    &    162               & \multicolumn{1}{c|}{}           &           \\ \hline
%Surveillance     & \multicolumn{1}{c|}{504}           & \multicolumn{1}{c|}{162} & \multicolumn{1}{c|}{}           &           \\ \hline
%\end{tabular}
%\end{center}
%\end{table}
%

%
\begin{table}[h]
\caption{Performance of the intent classifier in 2D and 3D  environments across 10 training-validation splits.}
\label{tab:accuracies_lstm_2d3d}
\begin{center}
\begin{tabular}{|c|cc|cc|}
\hline
& \multicolumn{2}{c|}{\begin{tabular}[c]{@{}c@{}}2D\\ environment\end{tabular}} & \multicolumn{2}{c|}{\begin{tabular}[c]{@{}c@{}}3D\\ environment\end{tabular}} \\ \hline
Accuracy($\%$) & \multicolumn{1}{c|}{mean}      & std.       & \multicolumn{1}{c|}{mean}      & std.                       \\ \hline
$W=50$   & \multicolumn{1}{c|}{$83.26$}       &    $0.85$        & \multicolumn{1}{c|}{$78.32$}       &   $1.03$                         \\ \hline
$W=150$    & \multicolumn{1}{c|}{$94.11$}   &    $1.31$       & \multicolumn{1}{c|}{$91.05$}   & $1.72$                           \\ \hline
\end{tabular}
\end{center}
\end{table}
%
%

%\begin{table}[h]
%\caption{Performance of the intent classifier in different environments.  }
%\label{tab:accuracies_lstm_2d3d}
%\begin{center}
%\begin{tabular}{|c|cc|cc|}
%\hline
%         & \multicolumn{2}{c|}{\begin{tabular}[c]{@{}c@{}}2D\\ environment\end{tabular}} & \multicolumn{2}{c|}{\begin{tabular}[c]{@{}c@{}}3D\\ environment\end{tabular}} \\ \hline
%Accuracy($\%$) & \multicolumn{1}{c|}{mean}      & std.       & \multicolumn{1}{c|}{mean}      & std.                       \\ \hline
%$W=50$   & \multicolumn{1}{c|}{$83.26$}       &    $0.05$        & \multicolumn{1}{c|}{$77.32$}       &   $0.03$                         \\ \hline
%$W=150$    & \multicolumn{1}{c|}{$95.11$}   &    $0.05$       & \multicolumn{1}{c|}{$94.64$}   & $0.002$                           \\ \hline
%\end{tabular}
%\end{center}
%\end{table}
%
\begin{figure}[htp]
    \centering
    \begin{tabular}{cc}
    \hspace{-0.15 in}
      \includegraphics[width=.5\linewidth]{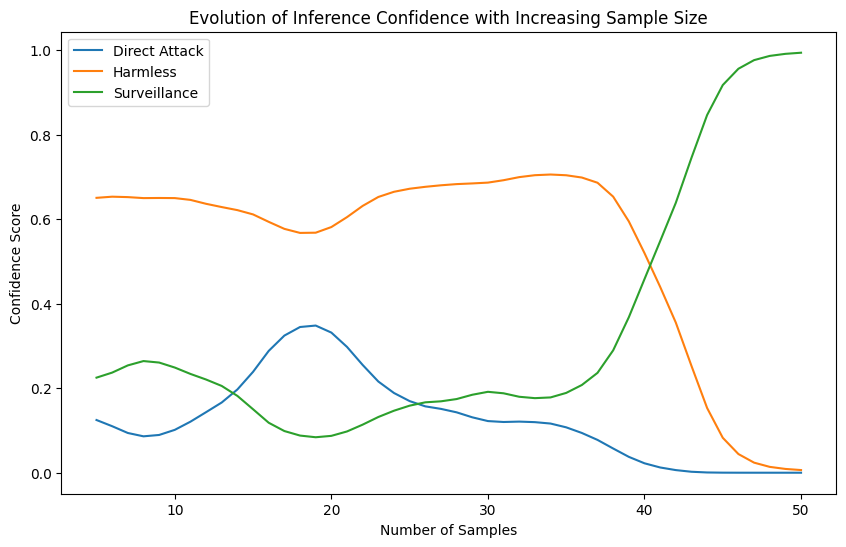} & 
      \hspace{-0.2 in}
      \includegraphics[width=.5\linewidth]{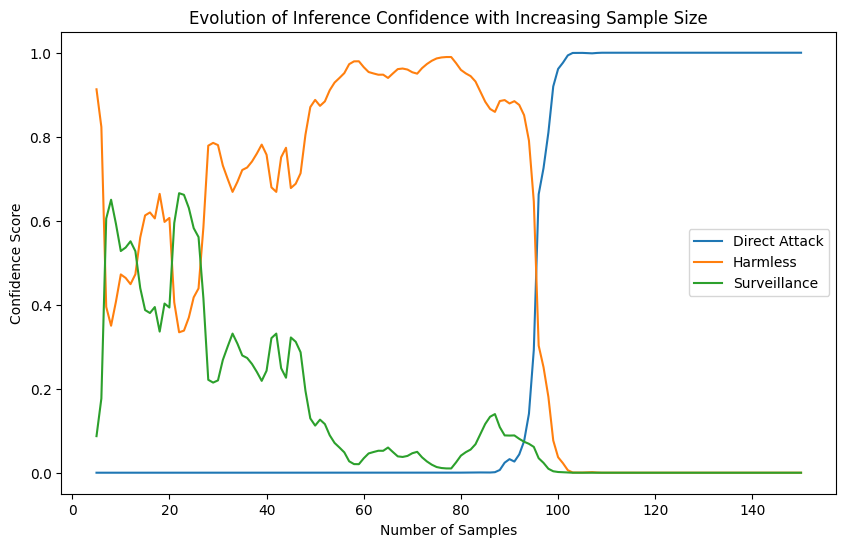}
    \end{tabular}
    \caption{{{Evolution of posterior probabilities for an example Direct Attack trajectory with number of time-samples, for classifiers trained using different feature extraction window sizes. \textit{[left]} $W=50,$ \textit{[right]} $W=150.$  In the left, the classifier becomes increasingly confident that the trajectory is Surveillance, which is not true. The classifier on the right becomes confident that it is Direct Attack after observing a larger number of samples. } }}
    \label{fig:sample_confidence_evolution}
\end{figure}

%\begin{table}[h]
%    \centering
%    \begin{tabular}{|c|c|c|}
%        \hline
%                 & 2D features  & 5D features \\ \hline
%       Full          &  & \\ \hline
%       Shortened     &  & \\ \hline
%    \end{tabular}
%    \caption{Experiment 2: Validation accuracy of Intent Recognition. Intents have different speed and acceleration profiles while path following.  }
%    \label{tab:accuracies_diff_speed_profiles}
%\end{table}

Table~\ref{tab:accuracies_lstm_2d3d} summarizes the performance of the neural network classifier for different cases. The feature extraction window $W$ impacts the accuracy in both 2D and 3D environments. Adjacent windows have a $90\%$ overlap. $W=50$  means under each window $50$ time samples were passed into the IMM filter for feature extraction which is later used by the network to compute the posterior probability of each intent. On an average, this needs a observation time (tracking after detection) of $5$ seconds. The simulations suggest that  with a sample rate of $10$ Hz, an observation time of $15$ seconds is needed to reach an intent classification accuracy of $95\%.$ For an accuracy of around $80\%,$ a least  $5$ seconds of observation is needed. Fig.~\ref{fig:sample_confidence_evolution} plots the evolution of confidence score of the classifier with time, in terms of posterior intent probabilities. Here, both the left and right plots correspond to the same trajectory but the classifier trained on $W=50$ features gives a different result from the classifier trained on $W=150$ features, as expected due to the different observation times.

%\kkComment{
%TO DO
%\begin{enumerate}
%    \item Generate trajectories and state sequences for each intent, and prepare labelled dataset.
%    \item Visualize some trajectories under each intent.
%    \item Define two neural network architectures for classification of intents.
%    \item Train the networks using the different loss functions.
%    \item Compare the performance using the metrics specified in the objectives.
%\end{enumerate}
%}

\section{Conclusion}

We presented a novel integrated framework for intent modelling and inference which can be applied for both autonomous and piloted aerial systems. We introduced a mathematical definition of intent using the concepts of critical waypoint pattern, critical waypoint regions and critical region ordered sets. An intent is also associated with motion process which describes motion between waypoints. This framework can be utilized for defining intent libraries and building intent inference systems as per  requirements, to protect a geo-fence. As future work we would like to conduct an elaborate simulation study that implements the intent inference framework for commercially available drones in realistic 3D environments using both manually defined and algorithmically derived critical regions for various various intents. Algorithms for deriving critical regions and critical waypoint patterns from experimental flight data is also an important direction for future work. Further, we would like to consider taking up the other objectives such as classification under time constraints and effective interception of unauthorized UAVs.

%\section*{Acknowledgement}
%This project has been supported by the National Research Council Canada's Grant No. ??  and the New Beginning Program. 

\typeout{}
\bibliographystyle{IEEEbib.bst}
\bibliography{intent.bib}

\end{document}